\documentclass[tradiabstract]{aa}

\usepackage{natbib}
\bibpunct{(}{)}{;}{a}{}{,}
\usepackage{graphicx}
\usepackage{txfonts}
\usepackage{epsfig}
\usepackage{longtable}
\usepackage{lscape}
\usepackage{amssymb}
\usepackage[colorlinks=true,citecolor=blue]{hyperref}
\usepackage{refcount}
\usepackage{txfonts}
\usepackage{epsfig}
\usepackage{longtable}
\usepackage{lscape}
\usepackage{amssymb} 
\usepackage[usenames,dvipsnames,svgnames,table]{xcolor}
\usepackage[colorlinks=true,citecolor=blue]{hyperref} 
\usepackage{refcount}

\def \inte {{\em INTEGRAL}}

\def \sw {{\em Swift}}
\def \xmm {{\em XMM--Newton}}

\def \hcm {\hbox {\ifmmode $ atom cm$^{-2}\else atom cm$^{-2}$\fi}}

\def \aip {AIP Conf. Ser.}
\def \ATel {ATel} 
\def \apj {ApJ}

\def \asr {AdSpR}

\def \gcn {GCN Circ.} 
\def \iaucirc {IAU Circ.}
\def \memsaits {Mem. Soc. Astron. Italiana Suppl.} 
 
\def \mnras {MNRAS}
\def \pasj {PASJ}
\def \ssr {SSRv}

\begin{document} 

\title{The 100-month \emph{Swift} Catalogue of Supergiant Fast X--ray  Transients}
\subtitle{II. SFXT diagnostics from outburst properties}

\author{P.\ Romano\inst{\ref{oab}}    
           \and
          P.A.\ Evans\inst{\ref{lu}}         
         \and
         E.\ Bozzo \inst{\ref{geneve},\ref{oar}}   
        \and 
          V.\ Mangano\inst{\ref{oab}}     
          \and
          S.\ Vercellone\inst{\ref{oab}}     
          \and
          C.\ Guidorzi\inst{\ref{unife},\ref{infn},\ref{oas}}   
         \and
         L.\ Ducci \inst{\ref{tubingen},\ref{geneve}}  
          \and  \\
          J.A.\ Kennea\inst{\ref{psu}} 
          \and 
          S.D.\ Barthelmy\inst{\ref{goddard}} 
          \and 
          D.M.\ Palmer\inst{\ref{lanl}}
         \and  
         H.A.\ Krimm\inst{\ref{nsf}}                      
         \and
         S.B.\ Cenko\inst{\ref{goddard}} 
 }  

\institute{
              INAF -- Osservatorio Astronomico di Brera, 
                  Via E.\ Bianchi 46, I-23807, Merate, Italy
                  \email{patrizia.romano@inaf.it}\label{oab}
              \and
              University of Leicester, X-ray and Observational Astronomy Group, 
                  School of Physics \& Astronomy, University Road, \\
                  Leicester, LE1 7RH \label{lu}
              \and 
             Department of Astronomy, Universit\'e de Gen\`eve, 16 chemin d'\'Ecogia, 1290 Versoix, Switzerland \label{geneve}
               \and
               INAF -- Osservatorio Astronomico di Roma, Via Frascati, 33, 00078 Monte Porzio Catone, Rome, Italy\label{oar}
           \and
            Department of Physics and Earth Science, University of Ferrara,
              Via Saragat 1, I-44122, Ferrara, Italy \label{unife}
           \and
             INFN -- Sezione di Ferrara, Via Saragat 1, I--44122, Ferrara, Italy \label{infn}
          \and
              INAF -- Osservatorio di Astrofisica e Scienza dello Spazio di Bologna, Via Piero Gobetti 93/3, I-40129 Bologna, Italy \label{oas}
            \and 
             Institut f\"ur Astronomie und Astrophysik, Kepler Center for Astro and Particle Physics, Universit\"at, 
                       T\"ubingen, Sand 1, 72076 T\"ubingen, Germany \label{tubingen}
            \and
              Department of Astronomy and Astrophysics, Pennsylvania State 
                  University, University Park, PA 16802, USA \label{psu}
            \and
                Astrophysics Science Division, NASA Goddard Space Flight Center, Greenbelt, MD 20771, USA \label{goddard}
                \and
                Los Alamos National Laboratory, B244, Los Alamos NM 87545, USA\label{lanl}
                \and 
                National Science Foundation, Alexandria, VA 22314, USA\label{nsf}
 }  

\date{Received 21 August 2022 /  Accepted 6 December 2022}

\abstract{Supergiant Fast X--ray Transients (SFXT) are High Mass X--ray Binaries (HMXBs) 
displaying X--ray outbursts that can reach peak luminosities up to 10$^{38}$~erg~s$^{-1}$ and spend 
most of their life in more quiescent states with luminosities as low as 10$^{32}$--10$^{33}$~erg~s$^{-1}$. 
During the quiescent states, less luminous flares are also frequently observed with luminosities 
of 10$^{34}$--10$^{35}$~erg~s$^{-1}$.
The main goal of the comprehensive and uniform analysis of the SFXT {\em Swift} triggers presented in this paper 
is to provide tools to predict whether a transient which has no known X-ray 
counterpart may be an SFXT candidate. These tools can be exploited for the development of 
future missions exploring the variable X-ray sky through large field-of-view instruments.
We examined all available data on outbursts of SFXTs that triggered the 
{\em Swift}/Burst Alert Telescope (BAT) collected between 2005 August 30 and 
2014 December 31, in particular those for which broad-band data, including the {\em Swift}/X-ray Telescope (XRT) 
ones, are also available. 
This work complements and extends our previous catalogue of SFXT flares detected by BAT  from 
2005 February 12 to 2013 May 31, since we now include 
the additional BAT triggers recorded until the end of 2014 (i.e.\ beyond the formal 
first 100 months of the {\em Swift} mission). 
Due to a change in the mission observational strategy, 
virtually no SFXT triggers obtained a broad-band response after 2014.  
We processed all BAT and XRT data uniformly by using the {\em Swift} Burst Analyser to 
produce spectral evolution dependent flux light curves for each outburst of the sample.  
The BAT data allowed us to infer useful diagnostics 
to set SFXT triggers apart from the general GRB population, showing that 
SFXTs give rise uniquely to image triggers and are simultaneously very long, 
faint, and `soft' hard-X-ray transients. 
We find that the BAT data alone can discriminate very well the SFXTs from 
other classes of fast transients such as anomalous X-ray pulsars and  soft gamma repeaters.
However, to distinguish SFXTs from, for instance,  accreting millisecond X-ray pulsars and  jetted tidal disruption events,
the XRT data collected around the time of the BAT triggers are shown to be decisive. 
The XRT observations of 35 (out of 52 in total) SFXT BAT triggers show that in the soft X-ray energy band, 
SFXTs display a decay in flux from the peak of the outburst 
of at least 3 orders of magnitude within a day and rarely undergo large re-brightening episodes, 
favouring in most cases a rapid decay down to the quiescent level within 3--5 days (at most). 
}


\maketitle

   \section{Introduction\label{sfxtcat2:intro}}

\begin{sidewaystable*}
\tabcolsep 3pt   	
 \caption{Sample of SFXTs and candidates: binary properties and optical counterparts.\label{sfxtcat2:tab:sampleproperties}}
 \centering
\tiny
 \begin{tabular}{l rrrrrrc ccccccc }
\hline 
\hline
\noalign{\smallskip} 
 Name                          & Spectral            &  Distance                                 &$P_{\rm spin}$                                          &$P_{\rm orb}$                                       &  $P_{\rm sup}$     & Eclipse & $e$  & Ref.\ &   \multicolumn{6}{c}{References}  \\
                                    & Type                 & (kpc)                                        &(s)	                                                    &(d)                                                   &(d)	                 &              &         &  Disc                                                                     & Sp.T.                                                                                & D                                                              &$P_{\rm spin}$         &$P_{\rm orb}$; $e$           &$P_{\rm sup}$      &  Eclipse      \\
(1) & (2) & (3) & (4) & (5) &  (6) &  (7) &  (8) &  (9) &   (10) & (11) & (12) & (13) & (14) & (15) \\ 
\hline  \noalign{\smallskip}
\object{IGR~J08408$-$4503}   &O8.5Ib-II(f)p       &$2.20^{+0.08}_{-0.09}$            &  --                                      &$9.5436\pm0.0002$                           &  285$\pm$10                             & N &  0.63$\pm$0.03  &\getrefnumber{tabGotz2006:08408-4503discovery}  &\getrefnumber{tabGOSSS2014}                                    & \getrefnumber{tabBailer-Jones2021:EDR3_distV}      &  --          &\getrefnumber{tabGamen2015:08408_period}; \getrefnumber{tabGamen2015:08408_period}       & \getrefnumber{tabGamen2015:08408_period}                & --                  \\ %
\object{IGR~J11215$-$5952}   &B0.5Ia                 &7.06$^{+0.55}_{-0.56}$                 &$186.78\pm0.3$               &$164.6$                                             &  --                              & N & $>0.8$             &\getrefnumber{tabLubinski2005}                                &\getrefnumber{tabVijapurkar1993}                               & \getrefnumber{tabBailer-Jones2021:EDR3_distV}       &\getrefnumber{tabSwank2007:atel999}     &\getrefnumber{tabRomano2009:11215_2008}; \getrefnumber{tabLorenzo2014:11215}     & --           & --     \\ %
\object{IGR~J16328$-$4726}   & O8Iafpe              &7.2$\pm0.3$                      &  --                                      &$10.076\pm0.003$                           &  --                              & N & --                       &\getrefnumber{tabBird2007:igr3cat}                           &\getrefnumber{tabColeiro2013:IR_IDs}                          &\getrefnumber{tabPersi2015:16328}       & --          &\getrefnumber{tabCorbet2010:16328-4726}                 & --                                                                               & --   \\ %
\object{IGR~J16418$-$4532}  &BN0.5Ia               &13                                       & $1209.12\pm0.42$          &$3.73886\pm0.00003$                     & $14.730\pm0.006$   & Y & --                       &\getrefnumber{tabTomsick2004:atel224}                   &\getrefnumber{tabRahoui2008}                                      &\getrefnumber{tabRahoui2008}                &\getrefnumber{tabDrave2013:16418}  &\getrefnumber{tabLevine2011} & \getrefnumber{tabCorbet2013:atel5126} &  \getrefnumber{tabCorbet2006:atel779}   \\ %
\object{IGR~J16465$-$4507}\tablefootmark{a}   &  B0.5Ib/O9.5Ia  &9.4/$9.5^{+14.1}_{-5.7}$      &$228\pm6$   &$30.243\pm0.035$                       &  --                               & N & $<0.6$,$<0.8$ &\getrefnumber{tabLutovinov2005}       &\getrefnumber{tabNegueruela2007},\getrefnumber{tabRahoui2008}    &\getrefnumber{tabRahoui2008},\getrefnumber{tabNespoli2008} & \getrefnumber{tabLutovinov2005}         &\getrefnumber{tabLaParola2010:16465-4507_period}; \getrefnumber{tabClark2010:16465-4507_period2}   & --   & --     \\ %
\object{IGR~J16479$-$4514}  &O8.5I                   &4.45                                     &  --                                      &$3.3193\pm0.0005$                          &  $11.880\pm0.002$ & Y  & --                      &\getrefnumber{tabMolkov2003:16479-4514}              &\getrefnumber{tabRahoui2008}                                     &\getrefnumber{tabArnason2021:GDR2_XRBs}                     &  --          &\getrefnumber{tabRomano2009:sfxts_paperV}   &\getrefnumber{tabCorbet2013:atel5126}    & \getrefnumber{tabBozzo2008:eclipse16479}          \\%
\object{XTE~J1739$-$302}	  &O8Iab(f)               &2.01$\pm$0.16                                     &  --                                     &$51.47\pm0.02$\tablefootmark{b}    &  --                               & N & $<0.8$            &\getrefnumber{tabSmith1997:iauc6748},\getrefnumber{tabSmith1998:17391-3021}   &\getrefnumber{tabNegueruela2006}   &\getrefnumber{tabBailer-Jones2021:EDR3_distV}  & --        &\getrefnumber{tabDrave2010:17391_3021_period}; \getrefnumber{tabDrave2010:17391_3021_period}      & -- & --       \\ %
\object{IGR~J17544$-$2619}  &O9Ib                    &3.0$\pm$0.2                      &$71.49\pm0.02$\tablefootmark{c}       &$4.926\pm0.001$                              &  --                              & N  &$>0$                &\getrefnumber{tabSunyaev2003}                                  &\getrefnumber{tabPellizza2006}                                   &\getrefnumber{tabGimenez-Garcia2016:17544}                       & \getrefnumber{tabDrave2012:17544_2619_pulsation}  &\getrefnumber{tabClark2009:17544-2619period}; \getrefnumber{tabClark2009:17544-2619period} & -- & -- \\%
\object{SAX~J1818.6$-$1703} &O9I-B1I              &$2.1\pm0.1$                      &  --                                      &$30\pm0.1$ 	                              &  --                               & N  &$0.3$--$0.4$ &\getrefnumber{tabzand1998}                                      &\getrefnumber{tabNegueruela2007:hmxbs}                   & \getrefnumber{tabTorrejon2010:hmxbs}        & --  &\getrefnumber{tabBird2009:sax1818.6_period},\getrefnumber{tabZurita2009:sax1818.6_period}; \getrefnumber{tabZurita2009:sax1818.6_period} & --  & --   \\ %
\object{AX~J1841.0$-$0536}  &B1Ib                    &$3.2_{-1.5}^{+2.0}$,$6.9\pm1.7$  &$4.7394\pm0.0008$?\tablefootmark{d}  & 6.4565$\pm$0.0055\tablefootmark{e}   &  --                   & N  &   --                   &\getrefnumber{tabBamba1999:iauc7324},\getrefnumber{tabBamba2001}  & \getrefnumber{tabNespoli2008}   & \getrefnumber{tabNespoli2008},\getrefnumber{tabSguera2009} &\getrefnumber{tabBamba2001}       & \getrefnumber{tabGonzalez2015:PhD}                                         & -- & --  \\ %
\object{AX~J1845.0$-$0433}  &O9.5I                  &6.4$\pm$0.76                      &  --                                    & $5.7195\pm0.0007$	                     &  --                                & N  &$<0.37$           &\getrefnumber{tabYamauchi1995:1845}                     &\getrefnumber{tabCoe1996:18450}                                &\getrefnumber{tabColeiro2013:distribHMXBs}                        & --      &\getrefnumber{tabGoossens2013:18450_period}; \getrefnumber{tabGoossens2013:18450_period}               & -- & --  \\ %
 \object{IGR~J18483$-$0311} 1  &B0.5Ia/B0-B1Iab &$2.83\pm0.05$                   &$21.0526\pm0.0005$\tablefootmark{f}   &$18.545\pm0.003$  &   --                               & N  &$0.4$               &\getrefnumber{tabChernyakova2003}    &\getrefnumber{tabRahoui2008},\getrefnumber{tabTorrejon2010:hmxbs}     &\getrefnumber{tabTorrejon2010:hmxbs}  &\getrefnumber{tabSguera2007}    & \getrefnumber{tabLevine2006:igr18483},\getrefnumber{tabLevine2011}; \getrefnumber{tabRomano2010:sfxts_18483}                              &--    & --   \\ %
 \noalign{\smallskip}
\object{2XMM~J185114.3}\tablefootmark{g}      &   --                       &12                                         & --                                     & --                                                         & --                                & --                       & --                                  & \getrefnumber{tabWatson2009:2XMMserendip},\getrefnumber{tabLin2012:XMMSerendip}                                 & --                                  &   \getrefnumber{tabRomano2016_3sfxt}  & --        &   -- & --                                  & --                                    \\ 
  \noalign{\smallskip}
  \hline
  \end{tabular}
\tablefoot{ 
\tablefoottext{a}{Classical HMXB, not considered further. }
\tablefoottext{b}{See \citet[][$P_{\rm orb}=12.8658\pm0.0073$\,d]{Romano2009:sfxts_paperV}.}
\tablefoottext{c}{See \citet[][$P_{\rm spin}=11.6\pm0.13$\,s]{Romano2015:17544sb}.}
\tablefoottext{d}{See \citet{Bozzo2011:18410}.}
\tablefoottext{e}{Tentative.}
\tablefoottext{f}{See \citet{Ducci2013:18483}.}
\tablefoottext{g}{Candidate SFXT.}
}
\tablebib{ 
\newcounter{ctr_tab1tabrefs} 
\setrefcountdefault{-99} 
\refstepcounter{ctr_tab1tabrefs}\label{tabGotz2006:08408-4503discovery}(\getrefnumber{tabGotz2006:08408-4503discovery})  \citet{Gotz2006:08408-4503discovery};  
\refstepcounter{ctr_tab1tabrefs}\label{tabGOSSS2014}(\getrefnumber{tabGOSSS2014})  \citet{GOSSS2014};  
\refstepcounter{ctr_tab1tabrefs}\label{tabBailer-Jones2021:EDR3_distV}(\getrefnumber{tabBailer-Jones2021:EDR3_distV})  \citet{Bailer-Jones2021:EDR3_distV};    
\refstepcounter{ctr_tab1tabrefs}\label{tabGamen2015:08408_period}(\getrefnumber{tabGamen2015:08408_period})  \citet{Gamen2015:08408_period};    
\refstepcounter{ctr_tab1tabrefs}\label{tabLubinski2005}(\getrefnumber{tabLubinski2005})  \citet{Lubinski2005};   
\refstepcounter{ctr_tab1tabrefs}\label{tabVijapurkar1993}(\getrefnumber{tabVijapurkar1993})  \citet{Vijapurkar1993};   
\refstepcounter{ctr_tab1tabrefs}\label{tabSwank2007:atel999}(\getrefnumber{tabSwank2007:atel999})  \citet{Swank2007:atel999};   
\refstepcounter{ctr_tab1tabrefs}\label{tabRomano2009:11215_2008}(\getrefnumber{tabRomano2009:11215_2008})  \citet{Romano2009:11215_2008};   
\refstepcounter{ctr_tab1tabrefs}\label{tabLorenzo2014:11215}(\getrefnumber{tabLorenzo2014:11215})  \citet{Lorenzo2014:11215};  
\refstepcounter{ctr_tab1tabrefs}\label{tabBird2007:igr3cat}(\getrefnumber{tabBird2007:igr3cat})  \citet{Bird2007:igr3cat};   
\refstepcounter{ctr_tab1tabrefs}\label{tabColeiro2013:IR_IDs}(\getrefnumber{tabColeiro2013:IR_IDs})  \citet{Coleiro2013:IR_IDs};  
\refstepcounter{ctr_tab1tabrefs}\label{tabPersi2015:16328}(\getrefnumber{tabPersi2015:16328})  \citet{Persi2015:16328};   
\refstepcounter{ctr_tab1tabrefs}\label{tabCorbet2010:16328-4726}(\getrefnumber{tabCorbet2010:16328-4726})  \citet{Corbet2010:16328-4726};   
\refstepcounter{ctr_tab1tabrefs}\label{tabTomsick2004:atel224}(\getrefnumber{tabTomsick2004:atel224})  \citet{Tomsick2004:atel224};   
\refstepcounter{ctr_tab1tabrefs}\label{tabRahoui2008}(\getrefnumber{tabRahoui2008})  \citet{Rahoui2008};  
\refstepcounter{ctr_tab1tabrefs}\label{tabDrave2013:16418}(\getrefnumber{tabDrave2013:16418})  \citet{Drave2013:16418};   
\refstepcounter{ctr_tab1tabrefs}\label{tabLevine2011}(\getrefnumber{tabLevine2011})  \citet{Levine2011};  
\refstepcounter{ctr_tab1tabrefs}\label{tabCorbet2013:atel5126}(\getrefnumber{tabCorbet2013:atel5126})  \citet{Corbet2013:atel5126};   
\refstepcounter{ctr_tab1tabrefs}\label{tabCorbet2006:atel779}(\getrefnumber{tabCorbet2006:atel779})  \citet{Corbet2006:atel779};  
\refstepcounter{ctr_tab1tabrefs}\label{tabLutovinov2004:16465-4507}(\getrefnumber{tabLutovinov2004:16465-4507})  \citet{Lutovinov2004:16465-4507};   
\refstepcounter{ctr_tab1tabrefs}\label{tabNegueruela2007}(\getrefnumber{tabNegueruela2007})  \citet{Negueruela2007};  
\refstepcounter{ctr_tab1tabrefs}\label{tabNespoli2008}(\getrefnumber{tabNespoli2008})  \citet{Nespoli2008};  
\refstepcounter{ctr_tab1tabrefs}\label{tabLutovinov2005}(\getrefnumber{tabLutovinov2005})  \citet{Lutovinov2005};     
\refstepcounter{ctr_tab1tabrefs}\label{tabLaParola2010:16465-4507_period}(\getrefnumber{tabLaParola2010:16465-4507_period})  \citet{LaParola2010:16465-4507_period};   
\refstepcounter{ctr_tab1tabrefs}\label{tabClark2010:16465-4507_period2}(\getrefnumber{tabClark2010:16465-4507_period2})  \citet{Clark2010:16465-4507_period2};   
\refstepcounter{ctr_tab1tabrefs}\label{tabMolkov2003:16479-4514}(\getrefnumber{tabMolkov2003:16479-4514})  \citet{Molkov2003:16479-4514};   
\refstepcounter{ctr_tab1tabrefs}\label{tabArnason2021:GDR2_XRBs}(\getrefnumber{tabArnason2021:GDR2_XRBs})  \citet{Arnason2021:GDR2_XRBs};   
\refstepcounter{ctr_tab1tabrefs}\label{tabRomano2009:sfxts_paperV}(\getrefnumber{tabRomano2009:sfxts_paperV})  \citet{Romano2009:sfxts_paperV};  
\refstepcounter{ctr_tab1tabrefs}\label{tabBozzo2008:eclipse16479}(\getrefnumber{tabBozzo2008:eclipse16479})  \citet{Bozzo2008:eclipse16479};   
\refstepcounter{ctr_tab1tabrefs}\label{tabSmith1997:iauc6748}(\getrefnumber{tabSmith1997:iauc6748})  \citet{Smith1997:iauc6748};  
\refstepcounter{ctr_tab1tabrefs}\label{tabSmith1998:17391-3021}(\getrefnumber{tabSmith1998:17391-3021})  \citet{Smith1998:17391-3021}; 
\refstepcounter{ctr_tab1tabrefs}\label{tabNegueruela2006}(\getrefnumber{tabNegueruela2006})  \citet{Negueruela2006}; 
\refstepcounter{ctr_tab1tabrefs}\label{tabDrave2010:17391_3021_period}(\getrefnumber{tabDrave2010:17391_3021_period})  \citet{Drave2010:17391_3021_period}; 
\refstepcounter{ctr_tab1tabrefs}\label{tabSunyaev2003}(\getrefnumber{tabSunyaev2003})  \citet{Sunyaev2003}; 
\refstepcounter{ctr_tab1tabrefs}\label{tabPellizza2006}(\getrefnumber{tabPellizza2006})  \citet{Pellizza2006}; 
\refstepcounter{ctr_tab1tabrefs}\label{tabGimenez-Garcia2016:17544}(\getrefnumber{tabGimenez-Garcia2016:17544})  \citet{Gimenez-Garcia2016:17544}; 
\refstepcounter{ctr_tab1tabrefs}\label{tabDrave2012:17544_2619_pulsation}(\getrefnumber{tabDrave2012:17544_2619_pulsation})  \citet{Drave2012:17544_2619_pulsation};  
\refstepcounter{ctr_tab1tabrefs}\label{tabClark2009:17544-2619period}(\getrefnumber{tabClark2009:17544-2619period})  \citet{Clark2009:17544-2619period}; 
\refstepcounter{ctr_tab1tabrefs}\label{tabzand1998}(\getrefnumber{tabzand1998})  \citet{zand1998}; 
\refstepcounter{ctr_tab1tabrefs}\label{tabNegueruela2007:hmxbs}(\getrefnumber{tabNegueruela2007:hmxbs})  \citet{Negueruela2007:hmxbs}; 
\refstepcounter{ctr_tab1tabrefs}\label{tabTorrejon2010:hmxbs}(\getrefnumber{tabTorrejon2010:hmxbs})  \citet{Torrejon2010:hmxbs}; 
\refstepcounter{ctr_tab1tabrefs}\label{tabBird2009:sax1818.6_period}(\getrefnumber{tabBird2009:sax1818.6_period})  \citet{Bird2009:sax1818.6_period}; 
\refstepcounter{ctr_tab1tabrefs}\label{tabZurita2009:sax1818.6_period}(\getrefnumber{tabZurita2009:sax1818.6_period})  \citet{Zurita2009:sax1818.6_period}; 
\refstepcounter{ctr_tab1tabrefs}\label{tabBamba1999:iauc7324}(\getrefnumber{tabBamba1999:iauc7324})  \citet{Bamba1999:iauc7324}; 
\refstepcounter{ctr_tab1tabrefs}\label{tabBamba2001}(\getrefnumber{tabBamba2001})  \citet{Bamba2001}; 
\refstepcounter{ctr_tab1tabrefs}\label{tabSguera2009}(\getrefnumber{tabSguera2009})  \citet{Sguera2009}; 
\refstepcounter{ctr_tab1tabrefs}\label{tabGonzalez2015:PhD}(\getrefnumber{tabGonzalez2015:PhD})  \citet{Gonzalez2015:PhD}; 
\refstepcounter{ctr_tab1tabrefs}\label{tabYamauchi1995:1845}(\getrefnumber{tabYamauchi1995:1845})  \citet{Yamauchi1995:1845}; 
\refstepcounter{ctr_tab1tabrefs}\label{tabCoe1996:18450}(\getrefnumber{tabCoe1996:18450})  \citet{Coe1996:18450};  
\refstepcounter{ctr_tab1tabrefs}\label{tabColeiro2013:distribHMXBs}(\getrefnumber{tabColeiro2013:distribHMXBs})  \citet{Coleiro2013:distribHMXBs};  
\refstepcounter{ctr_tab1tabrefs}\label{tabGoossens2013:18450_period}(\getrefnumber{tabGoossens2013:18450_period})  \citet{Goossens2013:18450_period}; 
\refstepcounter{ctr_tab1tabrefs}\label{tabChernyakova2003}(\getrefnumber{tabChernyakova2003})  \citet{Chernyakova2003};   
\refstepcounter{ctr_tab1tabrefs}\label{tabSguera2007}(\getrefnumber{tabSguera2007})  \citet{Sguera2007}; 
\refstepcounter{ctr_tab1tabrefs}\label{tabLevine2006:igr18483}(\getrefnumber{tabLevine2006:igr18483})  \citet{Levine2006:igr18483};  
\refstepcounter{ctr_tab1tabrefs}\label{tabRomano2010:sfxts_18483}(\getrefnumber{tabRomano2010:sfxts_18483})  \citet{Romano2010:sfxts_18483};  
\refstepcounter{ctr_tab1tabrefs}\label{tabWatson2009:2XMMserendip}(\getrefnumber{tabWatson2009:2XMMserendip})  \citet{Watson2009:2XMMserendip};  
\refstepcounter{ctr_tab1tabrefs}\label{tabLin2012:XMMSerendip}(\getrefnumber{tabLin2012:XMMSerendip})  \citet{Lin2012:XMMSerendip};  
\refstepcounter{ctr_tab1tabrefs}\label{tabRomano2016_3sfxt}(\getrefnumber{tabRomano2016_3sfxt})  \citet{Romano2016_3sfxt}.
}
\end{sidewaystable*}
 
Supergiant fast X--ray transients (SFXTs) are High Mass X--ray Binaries (HMXBs) associated with OB 
supergiants set apart from the classical supergiant HMXBs--which show luminosity variations by a factor of 10--50 on time scales 
of a few hundred to thousands of seconds--for a distinctive dynamic range up to 10$^5$ times larger 
\citep[SFXTs; see, e.g.,][for recent reviews]{Walter2015:HMXBs_IGR,nunez17}. 
With a quiescent level of $\sim 10^{32}$~erg\,s$^{-1}$ \citep[e.g.][]{zand2005,Bozzo2010:quiesc1739n08408} 
and peak luminosities up to 10$^{38}$~erg~s$^{-1}$  \citep[][]{Romano2015:17544sb}, 
the SFXT dynamic range can reach up to six orders of magnitude, 
despite SFXTs being  overall significantly sub-luminous with respect to classical supergiant HMXBs  
\citep[][]{Lutovinov2013:HMXBpop,Bozzo2015:underluminous}. 
SFXTs indeed go through rare outbursts that can last up to a few days \citep[as is the case of IGR~J11215$-$5952,][]{Romano2007} 
characterised by bright flares that typically last a few hours \citep[][]{Sguera2005, Romano2007, Romano2014:sfxts_catI,Paizis2014, sidoli18, sidoli19}, 
timescales that are significantly shorter than those observed in Be/X--ray binaries  \citep[see, e.g.,][for a review]{Reig2011}. 
Similarly to classical systems hosting accreting neutron stars (NS), the hard X--ray spectra during outbursts are power laws combined 
with high energy cut-offs, reminiscent of those of HMXBs, so it is generally assumed that all SFXTs might host a neutron star.  
Indeed, in about half of the SFXT sample (see Table~\ref{sfxtcat2:tab:sampleproperties}, Col.\ 4)  a pulsation is observed that is interpreted as a NS spin period 
and cyclotron resonant scattering features (CRSF)  have been proposed but never confirmed 
(see, \citealt{Sguera2010:18483} and \citealt{Ducci2013:18483} for IGR~J18483$-$0311;    
\citealt{Bhalerao2015:line17544} and \citealt{Bozzo2016:17544} for IGR~J17544$-$2619;  
\citealt{Sidoli2017:11215} and \citealt{Sidoli2020:11215} for IGR~J11215$-$5952).

The detailed physics behind the SFXT outbursts is still unknown, 
but it is probably related to either the properties of 
the wind from the supergiant companion  \citep{zand2005,Walter2007,Negueruela2008,Sidoli2007,Ducci2009,Ducci2010,Bozzo2021} 
or the compact object itself, a NS,  with the presence of mechanisms inhibiting accretion \citep[][]{Grebenev2007,Bozzo2008}. 
The most recent development \citep[][]{Shakura2012:quasi_spherical,Shakura2014:bright_flares} 
includes subsonic settling accretion regime combined with magnetic reconnections between the NS and the 
supergiant field transported by its wind \citep[see also][]{Hubrig2018}. 

The brightest flares from the SFXTs have been triggering the Burst Alert Telescope  \citep[BAT,][]{Barthelmy2005:BAT}   
on board the {\em Neil Gehrels Swift Observatory} \citep{Gehrels2004}  since early after launch. 
The \sw\ unique properties of automatic fast-slewing and broad-band energy coverage
have made it the sole observatory which can not only detect these events from the very beginning but also 
allow the follow-up of their evolution for days with a sensitive focusing X-ray instrument, 
the X--ray Telescope \citep[XRT,][]{Burrows2005:XRT}. 

In general, when a triggering event is recognised as originating from a 
previously known source (i.e.\ when it is in the on-board BAT catalogue), 
\sw\ does not slew to the target. 
For new or `special interest' sources, however,  \sw\ performs a slew, 
so that the narrow-field-instrument (NFI) can also image the target and provide broad band data. 
This has been the case for SFXTs since 2008 September 25 when, in order to 
ensure simultaneous NFI data, the \sw\ Team enabled automatic rapid slews 
to a pre-defined set of these objects following detection of flares by BAT 
as is routinely done for $\gamma$-ray bursts (GRBs). 
Furthermore, the triggering threshold for SFXTs was also lowered after each outburst, so as to allow further GRB-like response. 
This strategy has quickly tripled the available sets of data on SFXT outbursts and
allowed the arc-second localisation with the XRT 
of several confirmed SFXTs and a number of candidate sources in this class  
whose coordinates were only known 
at the arc-minute level \citep[e.g.][]{Kennea2005:16479-4514,Kennea2006:08408-4503,Romano2016_3sfxt}. 
This consequently helped the firm association of these objects with an optical/infrared counterpart. 
Further details and a review of the \sw\ SFXT Project can be found in \citet[][]{Romano2015:swift10}. 

Several bright flares were also independently caught by the 
BAT Transient Monitor\footnote{\href{http://swift.gsfc.nasa.gov/docs/swift/results/transients/}{http://swift.gsfc.nasa.gov/docs/swift/results/transients/. }}
\citep{Krimm2013:BATTM}, and observed with the XRT during the monitoring campaigns that were performed 
 on 4 SFXTs between 2007 and 2009 \citep[][and references therein]{Romano2011:sfxts_paperVI}. 

On 2012 May 14, the BAT triggering threshold was lowered from 6.4 to 5.8\,$\sigma$ in order to better hunt for 
short \citep[see, e.g.][]{Kouveliotou1993} GRBs. As a beneficial side effect, this also resulted in more detections of  
soft $\gamma$-ray repeaters (SGRs) and SFXTs.  
For the latter sources, this change allowed us to trigger on the faint flares of 
IGR~J16418$-$4532 on 2012 June 3, 2013 April 2, and 2015 April 27, 
and on fainter flares from IGR~J17544$-$2619  on 2012 July 24 and 2013 June 28, 
and XTE~J1739$-$302  on 2012 September 9 and 2012 November 11 
(see Table~\ref{sfxtcat2:tab:dataswift}). 
Starting from the end of 2014, the threshold for a slew for SFXTs was not lowered 
after each outburst; consequently, only increasingly brighter events were granted a GRB-like response, 
enabling broad band data collection. The triggering statistics changed dramatically and this resulted in virtually 
no additional SFXT triggers with broad band data after 2014. 

In a companion paper \citep[][Paper I]{Romano2014:sfxts_catI} we 
presented the catalogue of 1117 flares defined as detections in excess of 
5$\sigma$ registered by the BAT both on-board \sw\ and on the ground by the BAT Transient Monitor 
\citep{Krimm2013:BATTM},
independently if broad-band data were collected or not with the \sw\ NFI. 
These events were recorded from 11 SFXTs between 2005 February 12 and 2013 May 31. 
We showed that these flares/outbursts lasted just a few hundred seconds, achieving average 
fluxes as large as 100\,mCrab, (15--50\,keV) and average luminosities in the range  10$^{34}$--10$^{35}$~erg~s$^{-1}$.  
Based on our results, 
in combination with the first nine years of  \inte/ISGRI data \citep[][]{Paizis2014},  
\citet[][]{Ducci2014:sfxtN} derived the expected number of SFXTs emitting bright flares in the Milky Way,
$N\approx 37^{+53}_{-22}$, so that SFXTs may constitute a large portion of supergiant XRBs in the Galaxy.
 Given that the SFXT class currently includes only about a  dozen confirmed individuals, 
it is appropriate to devise methods to find the SFXTs still undetected.

In this paper (Paper II), we focus on bright SFXT flares that triggered the BAT and for which also complementary broad-band XRT data were simultaneously collected.   
The aim of our analysis is to exploit a uniform dataset of combined BAT+XRT data for its predictive potential.  
In Sect.~\ref{sfxtcat2:dataredu} we describe our samples and the uniform processing of the BAT and XRT data with the Burst Analyser 
\citep[][]{Evans2010:burstanalyser} to produce spectral evolution dependent flux light curves.  
In Sect.~\ref{sfxtcat2:results} we derive some characteristic properties of SFXTs as a sample both in the hard and in the soft X-ray band and 
we discuss our findings, considering future perspectives, thus tackling the issue of predicting whether a Galactic transient which has no 
known X-ray counterpart may be a SFXT candidate, for \sw\ and for other future missions exploring the X-ray variable sky 
with large field-of-view (FoV) instruments. We examine several diagnostics that help 
identifying SFXT candidates among the population of newly discovered fast X-ray transients.   
Finally, in Sect.~\ref{sfxtcat2:conclusions} we summarise our findings and draw our conclusions. 

 \setcounter{table}{1} 
\begin{table*}
 \tabcolsep 3pt         
\caption{The \sw{} SFXT outburst data. \label{sfxtcat2:tab:dataswift} }
\tiny
 \centering
  \begin{tabular}{lc  lrcccl cc}
\hline 
\hline 
\noalign{\smallskip} Name           & \multicolumn{5}{c}{Trigger}                                &\multicolumn{2}{l}{Data}                &\multicolumn{2}{c}{References}   \\
                        & N$^{\mathrm{a}}$    & \#$^{\mathrm{b}}$  & {\it S/N}$^{\mathrm{c}}$     &UT Date            &UT Time       &BAT ObsID          &XRT ObsID        &Discovery   & Refereed  \\
(1) & (2) & (3) & (4) & (5) &  (6) &  (7) &  (8) &  (9) &   (10) \\ 
\noalign{\smallskip}
\hline
\noalign{\smallskip} 
IGR~J08408$-$4503 &1 &232\,309  &8.08  &2006-10-04  &14:45:43  &00232309000 &00232309000,001  	                       &\getrefnumber{tabZiaeepour2006:gcn5687}           & \getrefnumber{tabGotz2007:08408-4503},\getrefnumber{tabRomano2009:sfxts_paper08408}  \\    
                                 &2 &316\,063  &7.38  &2008-07-05  &21:14:13  &00316063000 &00316063000,00030707003--012     &\getrefnumber{tabPalmer2008:gcn7946}                                                                           &  \getrefnumber{tabRomano2009:sfxts_paper08408}  \\       
                                 &3 &325\,461  &10.00 &2008-09-21  &07:55:09  &00325461000 &00325461000,00030707013--018    &\getrefnumber{tabMangano2008:gcn8279},\getrefnumber{tabMangano2008:atel1727}   & \getrefnumber{tabSidoli2009:sfxts_paperIV}   \\  
                                 &4 &361\,128$^{\mathrm{d}}$  &6.62  &2009-08-28  &22:51:47  &00361128000 &00030707019		                       &\getrefnumber{tabBarthelmy2009:atel2178}                                                                     & \getrefnumber{tabMangano2012:Gamma12}    \\       
                                 &5 &361\,129$^{\mathrm{d}}$  &10.26 &2009-08-28  &23:09:23  &00361129000 & ---			                               &\getrefnumber{tabBarthelmy2009:atel2178}                                                                     & \getrefnumber{tabMangano2012:Gamma12} \\           
                                 &6 &417\,420  &8.68  &2010-03-28  &15:53:38  &00417420000 &00417420000		                       &\getrefnumber{tabRomano2010:atel2520}                                                                        & \getrefnumber{tabMangano2012:Gamma12}  \\   
                                 &7 &501\,368  &7.28  &2011-08-25  &00:53:05  &00501368000 &00501368000,00037881002--011     &\getrefnumber{tabMangano2011:atel3586}                                                                       &\getrefnumber{tabMangano2012:Gamma12} \\   
                                 &8 &559\,642  &7.16  & 2013-07-02  &08:10:45 &00559642000 & 00559642000,00037881097--102  &\getrefnumber{tabRomano2013:atel5190}                                                                         & --- \\ 
\noalign{\smallskip}
 \hline
\noalign{\smallskip}
IGRJ~16328$-$4726 &1&354\,542  &7.69 &2009-06-10 &07:54:27          &00354542000  &00354542000--004                         & \getrefnumber{tabGrupe2009:16328-4726} & \getrefnumber{tabFiocchi2010:16328-4726},\getrefnumber{tabRomano2013:Cospar12} \\  
                                 &2&510\,701  &7.87 &2011-12-29 &06:39:20          &00510701000  & ---                                            & --- & \getrefnumber{tabRomano2014:sfxts_catI} \\  
  \noalign{\smallskip}
 \hline
\noalign{\smallskip}
IGR~J16418$-$4532
                                       &1&307\,208  &7.23   &2008-03-21 &23:01:33  & 00307208000   & ---                                                    & --- & \getrefnumber{tabRomano2014:sfxts_catI}  \\   
                                       &2&523\,489  &6.22   &2012-06-03 &18:08:48  & 00523489000   &00523489000--001                          &  \getrefnumber{tabRomano2012:atel4148} & \getrefnumber{tabRomano2013:MI50x_sfxts}  \\  
                                       &3&552\,677  &6.39   &2013-04-02 &11:56:30  &00552677000    &00552677000                                    &  \getrefnumber{tabRomano2013:atel4939} &  \getrefnumber{tabRomano2014:sfxts_catI}   \\  
                                       &4&571\,067  &7.26   &2013-09-17 &14:41:17  &00571067000    &00571067000,00032037008--014   & \getrefnumber{tabKrimm2013:atel5398}  &  ---   \\  
 \noalign{\smallskip}
\hline
\noalign{\smallskip} 
IGR~J16479$-$4514&1 &152\,652  &7.71  &2005-08-30  &04:08:48   &00152652000 &00152652000,00030296001--004     &\getrefnumber{tabKennea2005:16479-4514}       &\getrefnumber{tabKennea2006:transients},\getrefnumber{tabSidoli2008:sfxts_paperI},\getrefnumber{tabSguera2008}\\  
                                &2 &210\,886  &5.78  &2006-05-20  &17:32:39   &00210886000 & ---                                                 &\getrefnumber{tabMarkwardt2006:16479-4514}       & \getrefnumber{tabSidoli2008:sfxts_paperI} \\       
                                &3 &215\,914  &5.34  &2006-06-24  &20:19:59   &00215914000 & ---                                                 &\getrefnumber{tabSidoli2008:sfxts_paperI}       & \getrefnumber{tabSidoli2008:sfxts_paperI} \\    
                                &4 &286\,412  &9.98  &2007-07-29  &12:07:34   &00286412000 & ---                                                 &\getrefnumber{tabSidoli2008:sfxts_paperI}       & \getrefnumber{tabSidoli2008:sfxts_paperI} \\     
                                &5 &306\,829$^{\mathrm{e}}$  &12.02 &2008-03-19  &22:44:45   &00306829000 &00306829000,00030296029--033   &\getrefnumber{tabBarthelmy2008GCN7466},\getrefnumber{tabRomano2008:atel1435}  & \getrefnumber{tabRomano2008:sfxts_paperII}  \\     
                                &6 &306\,830$^{\mathrm{e}}$  &21.64  &2008-03-19  &22:59:57   &00306830000 & ---			                              &\getrefnumber{tabRomano2008:atel1435}       & \getrefnumber{tabRomano2008:sfxts_paperII}  \\     
                                &7 &312\,068  &7.21  &2008-05-21  &06:03:41   &00312068000 & ---			                              & ---                                                                  & \getrefnumber{tab2Romano2009:sfxts_paperV}    \\    
                                &8 &341\,452  &10.68  &2009-01-29  &06:33:08   &00341452000 &00341452000,00030296077--095   &\getrefnumber{tabRomano2009:atel1920},\getrefnumber{tabLaParola2009:atel1929}  & \getrefnumber{tabBozzo2009:16479-4514outburst},\getrefnumber{tab2Romano2009:sfxts_paperV}   \\     
                      & 9 & 599\,041  &17.22 &2014-05-15 &09:52:45   &00599041000 & ---                                           &\getrefnumber{tabRomano2014:GCN16268}   & --- \\         
\noalign{\smallskip}									 			    
\hline
\noalign{\smallskip} 
 XTE~J1739$-$302	&1 &282\,535  &6.53  &2007-06-18  &03:10:46  &00282535000 & ---			                                & ---                                                            & \getrefnumber{tabSidoli2008:sfxts_paperI}  \\          
                                &2 &308\,797  &7.83  &2008-04-08  &21:28:13  &00308797000 &00308797000		                        &\getrefnumber{tabRomano2008:atel1466} & \getrefnumber{tabSidoli2009:sfxts_paperIII}  \\   
                                &3 &319\,963$^{\mathrm{f}}$  &9.15  &2008-08-13  &23:49:17  &00319963000 &00319963000,00030987070--081       &\getrefnumber{tabRomano2008:atel1659} & \getrefnumber{tabSidoli2009:sfxts_paperIV}  \\      
                                &4 &319\,964$^{\mathrm{f}}$  &11.14 &2008-08-14 &00:12:53  &00319964000 & ---			                                &\getrefnumber{tabRomano2008:atel1659} & \getrefnumber{tabSidoli2009:sfxts_paperIV}  \\          
                               & 5 &346\,069  &6.81  &2009-03-10  &18:39:55  &00346069000 &00030987106--109                           &\getrefnumber{tabRomano2009:atel1961} & \getrefnumber{tabRomano2011:sfxts_paperVI}	\\       
                               & 6 &446\,475  &7.44  &2011-02-22  &07:21:37  &00446475000 &00446475000    	                               &\getrefnumber{tabRomano2011:atel3182} & \getrefnumber{tabFarinelli2012:sfxts_paperVIII} \\   
                               & 7 &533\,120  &5.81  &2012-09-09  &23:34:14  &00533120000 &00533120000,00030987192--195   &\getrefnumber{tabRomano2012:atel4366} &  \getrefnumber{tabRomano2014:sfxts_catI}	\\ 
                               & 8 &538\,084  &6.10  &2012-11-11  &09:35:02  &00538084000 &---                                                    &\getrefnumber{tabDelia2012:GCN13964} &  \getrefnumber{tabRomano2014:sfxts_catI}  \\  
\noalign{\smallskip}
\hline
\noalign{\smallskip} 
IGR~J17544$-$2619  
                  &        & BTM         & ---    &2007-11-08  &01:31:04  & ---             & ---                                                             &\getrefnumber{tabKrimm2007:ATel1265}      &  \getrefnumber{tabRomano2014:sfxts_catI}  \\   
                  &1      &308\,224 &9.10    &2008-03-31  &20:50:45  &00308224000    &00308224000,00035056021 	             &\getrefnumber{tabSidoli2008:atel1454}    & \getrefnumber{tabSidoli2009:sfxts_paperIII} \\    
                  &        &XRT          & ---    &2008-09-04  &00:19:00  & ---                       & 00035056061                                  &\getrefnumber{tabRomano2008:atel1697} & \getrefnumber{tabSidoli2009:sfxts_paperIV}   \\   
                  &        &BTM         &  ---   &2009-03-15  &23:52:40  & ---             & ---                                                            &\getrefnumber{tabKrimm2009:atel1971}    & \getrefnumber{tabRomano2011:sfxts_paperVI}    \\   
                  &2      &354\,221 &8.15   &2009-06-06  &07:48:59  &00354221000    &00354221000--1,00035056109--111 &\getrefnumber{tabRomano2009:atel2069} & \getrefnumber{tabRomano2011:sfxts_paperVI}  \\     
                  &3      &414\,875 &7.45   &2010-03-04  &23:13:54  &00414875000    &00414875000--1,0035056149             &\getrefnumber{tabRomano2010:atel2463} & \getrefnumber{tabRomano2011:sfxts_paperVII}    \\   
                  &4      &449\,907 &12.78 &2011-03-24  &01:56:57  &00449907000    &00449907000,00035056150--152       &\getrefnumber{tabRomano2011:atel3235} & \getrefnumber{tabFarinelli2012:sfxts_paperVIII}   \\    
                  &        &BTM          &  ---   &2012-04-12  &00:37:20  & ---                  & 00035056153--155                            &\getrefnumber{tabRomano2012:atel4040}   & \getrefnumber{tabRomano2014:sfxts_catI}     \\    
                  &5      &528\,432  &6.14   &2012-07-24  &04:52:47  & 00528432000   & 00528432000                                     &\getrefnumber{tabRomano2012:atel4275}   &  \getrefnumber{tabRomano2014:sfxts_catI}    \\  
                  &6      &559\,221  &6.02   & 2013-06-28 &07:26:21 & 00559221000   & 00559221000                                       & \getrefnumber{tabRomano2013:atel5179} & \getrefnumber{tabRomano2015:17544sb} \\ 
                  &7      &570\,402  &13.78 & 2013-09-11 &15:59:49  & 00570402000 & 00570402000                                        & \getrefnumber{tabRomano2013:atel5388} & \getrefnumber{tabRomano2015:17544sb} \\ 
                 &8 &599\,954$^{\mathrm{g}}$ & 9.23 & 2014-05-25 & 22:25:48 &00599954000  &00599954000,00035056156--160     & \getrefnumber{tabBarthelmy2014:GCN16330},\getrefnumber{tabRomano2014:atel6173} & ---  \\   
                 &9 &599\,955$^{\mathrm{g}}$ &  11.30 & 2014-05-25 & 22:33:08 &00599955000 &00599955000                                     & ---                                                              & --- \\ 
             &10   &614\,903 & 19.82 & 2014-10-10 & 15:04:19 &00614903000 &00614903000,00035056161--166     & \getrefnumber{tabBarthelmy2014:GCN16904},\getrefnumber{tabRomano2014:atel6566}  &  \getrefnumber{tabRomano2015:17544sb}  \\  
             &       &BTM        & ---    & 2015-02-24& 11:21:20 & ---                 &00035056167--168                             & \getrefnumber{tabKrimm2015:atel7137}  &  ---   \\   
 \noalign{\smallskip}
\hline
\noalign{\smallskip} 
SAX~J1818.6$-$1703&1&294\,385              &7.94  &2007-10-16  &04:14:30  &00294385000& ---                                                   & ---  & \getrefnumber{tabRomano2014:sfxts_catI}                               \\    
                                  &2&306\,379              &8.02  &2008-03-15  &15:49:01  &00306379000& ---                                                   &\getrefnumber{tabBarthelmy2008GCN7419} &   \getrefnumber{tabRomano2014:sfxts_catI} \\    
                                  &3&351\,323              &7.12  &2009-05-06  &14:02:11  &00351323000& 00351323000,00031409001--010    &\getrefnumber{tabRomano2009:atel2044} & \getrefnumber{tabSidoli2009:sfxts_sax1818} \\    
                                  &4&361\,958              &6.63  &2009-09-05  &11:15:15  &00361958000& 00361958000	 	                          &\getrefnumber{tabRomano2009:atel2191} &  \getrefnumber{tabRomano2012:probes11}  \\  
                                  &5&374\,869              &7.34  &2009-11-04  &07:24:11  &00374869000& ---                                                    &\getrefnumber{tabRomano2009:atel2279} &  \getrefnumber{tabRomano2014:sfxts_catI}     \\    
                               & 6 & 591\,551       & 6.16 &2014-03-13  &08:28:20  &00591551000& 00591551000                                    & \getrefnumber{tabKennea2014:atel5980} &  ---     \\    
\noalign{\smallskip}
\hline
\noalign{\smallskip}
AX~J1841.0$-$0536 
                   &1&423\,958 &6.87  &2010-06-05  &17:23:30  &00423958000 & 00423958000--1,00030988093--101   &\getrefnumber{tabDePasquale2010:atel2661},\getrefnumber{tabRomano2010:atel2662}  & \getrefnumber{tabRomano2011:sfxts_paperVII}         \\  
                   &2&455\,967 &6.91  &2011-06-24  &14:27:05  &00455967000 & 00030988107--114                                &\getrefnumber{tabMangano2011:atel3453}    &   \getrefnumber{tabRomano2014:sfxts_catI}  \\  
                   &3&524\,364 &8.47  &2012-06-14  &19:11:52  &00524364000 & 00524364000,00030988115--120         &\getrefnumber{tabRomano2012:atel4176}      & \getrefnumber{tabRomano2013:MI50x_sfxts} \\ 
                   &4&528\,411 &9.81  &2012-07-24  &00:40:15  &00528411000 & 00528411000                                          & \getrefnumber{tabRomano2012:atel4276}     &  \getrefnumber{tabRomano2014:sfxts_catI}  \\ 
\noalign{\smallskip}
\hline
\noalign{\smallskip} 
AX~J1845.0$-$0433 
                 &1&162\,526 &5.43  &2005-11-04 &22:26:48  &00162526000  & ---                                                         & ---                                                           & \getrefnumber{tabRomano2014:sfxts_catI}			     \\    
                 &2&355\,911 &6.99  &2009-06-28 &08:06:03  &00355911000  &00355911000                                          &\getrefnumber{tabRomano2009:atel2102} &  \getrefnumber{tabRomano2014:sfxts_catI} \\   
                 &3&521\,567 &7.26  &2012-05-05 &01:44:40  &00521567000  &00521567000--001,00032456001--023   &\getrefnumber{tabRomano2012:atel4095} &  \getrefnumber{tabRomano2013:Cospar12} \\   
\noalign{\smallskip}
\hline
\noalign{\smallskip} 
IGR~J18483$-$0311 &1&321\,750 &11.48  &2008-08-24 &10:48:29  &00321750000 & ---                                        & ---                                                          & \getrefnumber{tabRomano2014:sfxts_catI}	\\  
                                 &&BTM &  ---  &2011-11-23 &                & ---                    & ---                                               & \getrefnumber{tabKrimm2011:atel3780} & \getrefnumber{tabRomano2014:sfxts_catI}	\\  
\noalign{\smallskip}
\hline
\noalign{\smallskip} 
 2XMM~J185114.3&1 &524\,542  &7.11  &2012-06-17  &15:46:56  &00524542000 &00524542000--005,007--017  	       &\getrefnumber{tabBarthelmy2012:GCN13367} & \getrefnumber{tabRomano2016_3sfxt_t2} \\       
 \noalign{\smallskip}
\hline
  \end{tabular}
\tablefoot{
\tablefoottext{a}{Progressive number of BAT trigger.} 
\tablefoottext{b}{BAT Trigger number. }
\tablefoottext{c}{On-board significance of detections of BAT trigger in units of $\sigma$.} 
\tablefoottext{d-g}{The source triggered the BAT twice within a few minutes.} 
}
\end{table*}

\setcounter{table}{1}  
\begin{table*}
 \tabcolsep 3pt   	
\caption{continued.}
 \centering
  \begin{tabular}{llcllllll}
\hline
\end{tabular}
\tablebib{
\newcounter{ctr_tab2tabrefs} 
\setrefcountdefault{-99} 
\refstepcounter{ctr_tab2tabrefs}\label{tabZiaeepour2006:gcn5687}(\getrefnumber{tabZiaeepour2006:gcn5687}) \citet{Ziaeepour2006:gcn5687}; 
\refstepcounter{ctr_tab2tabrefs}\label{tabGotz2007:08408-4503}(\getrefnumber{tabGotz2007:08408-4503}) \citet{Gotz2007:08408-4503}; 
\refstepcounter{ctr_tab2tabrefs}\label{tabRomano2009:sfxts_paper08408}(\getrefnumber{tabRomano2009:sfxts_paper08408}) \citet{Romano2009:sfxts_paper08408};
\refstepcounter{ctr_tab2tabrefs}\label{tabPalmer2008:gcn7946}(\getrefnumber{tabPalmer2008:gcn7946}) \citet{Palmer2008:gcn7946}; 
\refstepcounter{ctr_tab2tabrefs}\label{tabMangano2008:gcn8279}(\getrefnumber{tabMangano2008:gcn8279}) \citet{Mangano2008:gcn8279};
\refstepcounter{ctr_tab2tabrefs}\label{tabMangano2008:atel1727}(\getrefnumber{tabMangano2008:atel1727}) \citet{Mangano2008:atel1727};
\refstepcounter{ctr_tab2tabrefs}\label{tabSidoli2009:sfxts_paperIV}(\getrefnumber{tabSidoli2009:sfxts_paperIV}) \citet{Sidoli2009:sfxts_paperIV};
\refstepcounter{ctr_tab2tabrefs}\label{tabBarthelmy2009:atel2178}(\getrefnumber{tabBarthelmy2009:atel2178})  \citet{Barthelmy2009:atel2178};
\refstepcounter{ctr_tab2tabrefs}\label{tabMangano2012:Gamma12}(\getrefnumber{tabMangano2012:Gamma12})  \citet{Mangano2012:Gamma12}; 
\refstepcounter{ctr_tab2tabrefs}\label{tabRomano2010:atel2520}(\getrefnumber{tabRomano2010:atel2520})  \citet{Romano2010:atel2520};
\refstepcounter{ctr_tab2tabrefs}\label{tabMangano2011:atel3586}(\getrefnumber{tabMangano2011:atel3586})  \citet{Mangano2011:atel3586};
\refstepcounter{ctr_tab2tabrefs}\label{tabRomano2013:atel5190}(\getrefnumber{tabRomano2013:atel5190})  \citet{Romano2013:atel5190};
\refstepcounter{ctr_tab2tabrefs}\label{tabGrupe2009:16328-4726}(\getrefnumber{tabGrupe2009:16328-4726})  \citet{Grupe2009:16328-4726};
\refstepcounter{ctr_tab2tabrefs}\label{tabFiocchi2010:16328-4726}(\getrefnumber{tabFiocchi2010:16328-4726})  \citet{Fiocchi2010:16328-4726};
\refstepcounter{ctr_tab2tabrefs}\label{tabRomano2013:Cospar12}(\getrefnumber{tabRomano2013:Cospar12})  \citet{Romano2013:Cospar12};
\refstepcounter{ctr_tab2tabrefs}\label{tabRomano2014:sfxts_catI}(\getrefnumber{tabRomano2014:sfxts_catI})  \citet{Romano2014:sfxts_catI};
\refstepcounter{ctr_tab2tabrefs}\label{tabRomano2012:atel4148}(\getrefnumber{tabRomano2012:atel4148})  \citet{Romano2012:atel4148}; 
\refstepcounter{ctr_tab2tabrefs}\label{tabRomano2013:MI50x_sfxts} (\getrefnumber{tabRomano2013:MI50x_sfxts})  \citet{Romano2013:MI50x_sfxts}; 
\refstepcounter{ctr_tab2tabrefs}\label{tabRomano2013:atel4939}(\getrefnumber{tabRomano2013:atel4939})  \citet{Romano2013:atel4939};
\refstepcounter{ctr_tab2tabrefs}\label{tabKrimm2013:atel5398}(\getrefnumber{tabKrimm2013:atel5398})  \citet{Krimm2013:atel5398};
\refstepcounter{ctr_tab2tabrefs}\label{tabKennea2005:16479-4514}(\getrefnumber{tabKennea2005:16479-4514})  \citet{Kennea2005:16479-4514};
\refstepcounter{ctr_tab2tabrefs}\label{tabKennea2006:transients}(\getrefnumber{tabKennea2006:transients})  \citet{Kennea2006:transients};
\refstepcounter{ctr_tab2tabrefs}\label{tabSidoli2008:sfxts_paperI}(\getrefnumber{tabSidoli2008:sfxts_paperI})  \citet{Sidoli2008:sfxts_paperI};
\refstepcounter{ctr_tab2tabrefs}\label{tabSguera2008}(\getrefnumber{tabSguera2008})  \citet{Sguera2008};
\refstepcounter{ctr_tab2tabrefs}\label{tabMarkwardt2006:16479-4514}(\getrefnumber{tabMarkwardt2006:16479-4514})  \citet{Markwardt2006:16479-4514};
\refstepcounter{ctr_tab2tabrefs}\label{tabBarthelmy2008GCN7466}(\getrefnumber{tabBarthelmy2008GCN7466})  \citet{Barthelmy2008GCN7466};
\refstepcounter{ctr_tab2tabrefs}\label{tabRomano2008:atel1435}(\getrefnumber{tabRomano2008:atel1435})  \citet{Romano2008:atel1435};
\refstepcounter{ctr_tab2tabrefs}\label{tabRomano2008:sfxts_paperII}(\getrefnumber{tabRomano2008:sfxts_paperII})  \citet{Romano2008:sfxts_paperII};
\refstepcounter{ctr_tab2tabrefs}\label{tab2Romano2009:sfxts_paperV}(\getrefnumber{tab2Romano2009:sfxts_paperV})  \citet{Romano2009:sfxts_paperV};
\refstepcounter{ctr_tab2tabrefs}\label{tabRomano2009:atel1920}(\getrefnumber{tabRomano2009:atel1920})  \citet{Romano2009:atel1920};
\refstepcounter{ctr_tab2tabrefs}\label{tabLaParola2009:atel1929}(\getrefnumber{tabLaParola2009:atel1929})  \citet{LaParola2009:atel1929};
\refstepcounter{ctr_tab2tabrefs}\label{tabBozzo2009:16479-4514outburst}(\getrefnumber{tabBozzo2009:16479-4514outburst})  \citet{Bozzo2009:16479-4514outburst};
\refstepcounter{ctr_tab2tabrefs}\label{tabRomano2014:GCN16268}(\getrefnumber{tabRomano2014:GCN16268})  \citet{Romano2014:GCN16268};
\refstepcounter{ctr_tab2tabrefs}\label{tabRomano2008:atel1466}(\getrefnumber{tabRomano2008:atel1466})  \citet{Romano2008:atel1466};
\refstepcounter{ctr_tab2tabrefs}\label{tabSidoli2009:sfxts_paperIII}(\getrefnumber{tabSidoli2009:sfxts_paperIII})  \citet{Sidoli2009:sfxts_paperIII};
\refstepcounter{ctr_tab2tabrefs}\label{tabRomano2008:atel1659}(\getrefnumber{tabRomano2008:atel1659})  \citet{Romano2008:atel1659};
\refstepcounter{ctr_tab2tabrefs}\label{tabRomano2009:atel1961}(\getrefnumber{tabRomano2009:atel1961})  \citet{Romano2009:atel1961};
\refstepcounter{ctr_tab2tabrefs}\label{tabRomano2011:sfxts_paperVI}(\getrefnumber{tabRomano2011:sfxts_paperVI})  \citet{Romano2011:sfxts_paperVI};
\refstepcounter{ctr_tab2tabrefs}\label{tabRomano2011:atel3182}(\getrefnumber{tabRomano2011:atel3182})  \citet{Romano2011:atel3182};
\refstepcounter{ctr_tab2tabrefs}\label{tabFarinelli2012:sfxts_paperVIII}(\getrefnumber{tabFarinelli2012:sfxts_paperVIII})  \citet{Farinelli2012:sfxts_paperVIII};
\refstepcounter{ctr_tab2tabrefs}\label{tabRomano2012:atel4366}(\getrefnumber{tabRomano2012:atel4366})  \citet{Romano2012:atel4366};
\refstepcounter{ctr_tab2tabrefs}\label{tabDelia2012:GCN13964}(\getrefnumber{tabDelia2012:GCN13964})  \citet{Delia2012:GCN13964};
\refstepcounter{ctr_tab2tabrefs}\label{tabKrimm2007:ATel1265}(\getrefnumber{tabKrimm2007:ATel1265})  \citet{Krimm2007:ATel1265};
 \refstepcounter{ctr_tab2tabrefs}\label{tabSidoli2008:atel1454}(\getrefnumber{tabSidoli2008:atel1454})  \citet{Sidoli2008:atel1454}; 
 \refstepcounter{ctr_tab2tabrefs}\label{tabRomano2008:atel1697}(\getrefnumber{tabRomano2008:atel1697})  \citet{Romano2008:atel1697}; 
 \refstepcounter{ctr_tab2tabrefs}\label{tabKrimm2009:atel1971}(\getrefnumber{tabKrimm2009:atel1971})  \citet{Krimm2009:atel1971}; 
 \refstepcounter{ctr_tab2tabrefs}\label{tabRomano2009:atel2069}(\getrefnumber{tabRomano2009:atel2069})  \citet{Romano2009:atel2069}; 
 \refstepcounter{ctr_tab2tabrefs}\label{tabRomano2010:atel2463}(\getrefnumber{tabRomano2010:atel2463})  \citet{Romano2010:atel2463};
 \refstepcounter{ctr_tab2tabrefs}\label{tabRomano2011:sfxts_paperVII}(\getrefnumber{tabRomano2011:sfxts_paperVII})  \citet{Romano2011:sfxts_paperVII}; 
 \refstepcounter{ctr_tab2tabrefs}\label{tabRomano2011:atel3235}(\getrefnumber{tabRomano2011:atel3235})  \citet{Romano2011:atel3235};
\refstepcounter{ctr_tab2tabrefs}\label{tabRomano2012:atel4040}(\getrefnumber{tabRomano2012:atel4040})  \citet{Romano2012:atel4040}; 
 \refstepcounter{ctr_tab2tabrefs}\label{tabRomano2012:atel4275}(\getrefnumber{tabRomano2012:atel4275})  \citet{Romano2012:atel4275}; 
 \refstepcounter{ctr_tab2tabrefs}\label{tabRomano2013:atel5179}(\getrefnumber{tabRomano2013:atel5179})  \citet{Romano2013:atel5179}; 
 \refstepcounter{ctr_tab2tabrefs}\label{tabRomano2015:17544sb}(\getrefnumber{tabRomano2015:17544sb})  \citet{Romano2015:17544sb}; 
 \refstepcounter{ctr_tab2tabrefs}\label{tabRomano2013:atel5388}(\getrefnumber{tabRomano2013:atel5388})  \citet{Romano2013:atel5388} ;
 \refstepcounter{ctr_tab2tabrefs}\label{tabBarthelmy2014:GCN16330}(\getrefnumber{tabBarthelmy2014:GCN16330})  \citet{Barthelmy2014:GCN16330}; 
 \refstepcounter{ctr_tab2tabrefs}\label{tabRomano2014:atel6173}(\getrefnumber{tabRomano2014:atel6173})  \citet{Romano2014:atel6173}; 
 \refstepcounter{ctr_tab2tabrefs}\label{tabBarthelmy2014:GCN16904}(\getrefnumber{tabBarthelmy2014:GCN16904})  \citet{Barthelmy2014:GCN16904}; 
 \refstepcounter{ctr_tab2tabrefs}\label{tabRomano2014:atel6566}(\getrefnumber{tabRomano2014:atel6566})  \citet{Romano2014:atel6566}; 
\refstepcounter{ctr_tab2tabrefs}\label{tabKrimm2015:atel7137}(\getrefnumber{tabKrimm2015:atel7137})  \citet{Krimm2015:atel7137};  
\refstepcounter{ctr_tab2tabrefs}\label{tabBarthelmy2008GCN7419}(\getrefnumber{tabBarthelmy2008GCN7419})  \citet{Barthelmy2008GCN7419};
\refstepcounter{ctr_tab2tabrefs}\label{tabRomano2009:atel2044}(\getrefnumber{tabRomano2009:atel2044})  \citet{Romano2009:atel2044};
\refstepcounter{ctr_tab2tabrefs}\label{tabSidoli2009:sfxts_sax1818}(\getrefnumber{tabSidoli2009:sfxts_sax1818})  \citet{Sidoli2009:sfxts_sax1818};
\refstepcounter{ctr_tab2tabrefs}\label{tabRomano2009:atel2191}(\getrefnumber{tabRomano2009:atel2191})  \citet{Romano2009:atel2191};
\refstepcounter{ctr_tab2tabrefs}\label{tabRomano2012:probes11}(\getrefnumber{tabRomano2012:probes11})  \citet{Romano2012:probes11};
\refstepcounter{ctr_tab2tabrefs}\label{tabRomano2009:atel2279}(\getrefnumber{tabRomano2009:atel2279})  \citet{Romano2009:atel2279};
\refstepcounter{ctr_tab2tabrefs}\label{tabKennea2014:atel5980}(\getrefnumber{tabKennea2014:atel5980})  \citet{Kennea2014:atel5980};
\refstepcounter{ctr_tab2tabrefs}\label{tabDePasquale2010:atel2661}(\getrefnumber{tabDePasquale2010:atel2661})  \citet{DePasquale2010:atel2661};
\refstepcounter{ctr_tab2tabrefs}\label{tabRomano2010:atel2662}(\getrefnumber{tabRomano2010:atel2662})  \citet{Romano2010:atel2662};
\refstepcounter{ctr_tab2tabrefs}\label{tabMangano2011:atel3453}(\getrefnumber{tabMangano2011:atel3453})  \citet{Mangano2011:atel3453};
\refstepcounter{ctr_tab2tabrefs}\label{tabRomano2012:atel4176}(\getrefnumber{tabRomano2012:atel4176})  \citet{Romano2012:atel4176};
\refstepcounter{ctr_tab2tabrefs}\label{tabRomano2012:atel4276}(\getrefnumber{tabRomano2012:atel4276})  \citet{Romano2012:atel4276};
\refstepcounter{ctr_tab2tabrefs}\label{tabRomano2009:atel2102}(\getrefnumber{tabRomano2009:atel2102})  \citet{Romano2009:atel2102};
\refstepcounter{ctr_tab2tabrefs}\label{tabRomano2012:atel4095}(\getrefnumber{tabRomano2012:atel4095})  \citet{Romano2012:atel4095};
\refstepcounter{ctr_tab2tabrefs}\label{tabKrimm2011:atel3780}(\getrefnumber{tabKrimm2011:atel3780})  \citet{Krimm2011:atel3780};
\refstepcounter{ctr_tab2tabrefs}\label{tabBarthelmy2012:GCN13367}(\getrefnumber{tabBarthelmy2012:GCN13367})  \citet{Barthelmy2012:GCN13367};
\refstepcounter{ctr_tab2tabrefs}\label{tabRomano2016_3sfxt_t2}(\getrefnumber{tabRomano2016_3sfxt_t2})  \citet{Romano2016_3sfxt}.
}
\end{table*}

   \section{Data sample, reduction and analysis} \label{sfxtcat2:dataredu}

   \subsection{The SFXT sample} \label{sfxtcat2:sfxtsample}
 
Our sample of SFXTs was selected from the literature based on reports of bright flares 
(peak $L\ga10^{36}$ erg s$^{-1}$) recorded by ASCA, RXTE, \inte, and \sw\ 
(their properties were detailed in Paper I, Sect.~2).  
It is convenient, given the different timescales and dynamic range on which the SFXT activity is observed,
to recapitulate our distinction between a flare, which is a state of enhanced emission generally lasting 
for a few hours (average luminosities in the range  10$^{34}$--10$^{35}$~erg\,s$^{-1}$, Paper I)  
and an outburst, which is composed of several very bright flares (those reaching luminosities up to 
$\sim 10^{38}$\,erg\,s$^{-1}$) that  can last, depending on the source, up to a day or longer.
Let us also briefly re-state that our operative definition of confirmed and candidate SFXTs 
is based on the availability (or lack) of an optical classification of the companion,
the former being a transient that has shown repeated, 
large dynamical range, flaring/outbursting activity and is  firmly associated with an OB supergiant,  
the latter being a transient that has shown similar X--ray behaviour 
but has no confirmed association with an OB supergiant companion.  

Here, Table~\ref{sfxtcat2:tab:sampleproperties} lists, for all SFXTs considered in Paper I, the 
information updated according to the most recent literature, on spectral type and distance (Col.\ 2, 3), 
the spin, orbital, and super-orbital periods  (Col.\ 4, 5, and 6), 
the presence of eclipses in the X--ray light curve (Col.\ 7), orbital eccentricities (Col.\ 8),  and 
the references to discovery, and counterpart properties (Col.\ 9--15).  
Differently from Table~3 of Paper I, IGR~J16465$-$4507 (which never triggered the BAT)
was found to be a classical HMXB and not an SFXT \citep{Romano2014:sfxts_paperX}, 
while the source 2XMM~J185114.3$-$000004 has been included in our sample 
as a promising SFXT candidate \citep[][]{Romano2016_3sfxt}. 
We note that IGR~J11215$-$5952 also never triggered the BAT so  
it is not considered in this work.

Differently from Paper I, the sample of outbursts actually exceeds the first 100 months 
of the \sw\ mission, and includes all those recorded until the end of 2014  
(thus including 3 triggers that occurred after 2013 May 31, and 5 that occurred during 2014).  
In Table~\ref{sfxtcat2:tab:dataswift} we summarise the \sw\ observations of the 11 
sources that triggered the BAT or for which outbursts where observed in the X--ray 
with XRT during our monitoring campaigns. 
We report the progressive number of triggers (Col.\ 2), 
the BAT trigger number  (Col.\ 3, 
             where XRT and BTM stand for XRT and BAT Transient Monitor triggers, respectively),
the BAT trigger significance of detection  (Col.\ 4), 
             trigger UT dates (Col.\ 5 and 6), 
the kind of data available for each outburst (Col.\ 7 and 8),
the references to the discovery  
and the in-depth outburst analysis (Col.\ 9 and 10), when already published. 
We also note the instances (4) where subsequent bright flares within the same outburst triggered the BAT.

From launch to the end of December 2014, {\it Swift}/BAT has detected 
52 outbursts (for a total of 56 on-board triggers). 
Of these, 35 have broad band coverage, the great majority of which (27) 
thanks to the SFXTs being added to the BAT special interest source 
list\footnote{For completeness, 
we note that two SFXT outbursts were also detected from  IGR~J16418$-$4532 after the end of 2014: 
the 2015 April 27 one \citep[trigger 639\,199,][]{Barthelmy2015:GCN17764,Romano2015:atel7454}
which, based on the ground-based analysis revealed a significance much smaller than the one derived on-board, 
so that XRT only observed the normal out-of-outburst activity for this source, 
and the 2021 September 18 one \citep[trigger 1\,073\,821,][]{Sbarufatti2021:atel14924}. 
The latter event will be considered separately in a forthcoming publication. 
}.

   \subsection{The GRB sample} \label{sfxtcat2:grbsample}

GRBs are the main triggering sources for the BAT, so they make  
a natural control sample for the SFXT one. 
For this work we considered all GRBs that triggered the BAT until end of 2014 (839 triggers).
 The data from \sw-detected GRBs are regularly processed with the \sw\ 
Burst Analyser 
\citep{Evans2010:burstanalyser}, which ensures uniformity in treatment with SFXTs.

   \subsection{Other transients} \label{sfxtcat2:transsample}

It is also worthwhile to consider other transients, both Galactic and extragalactic,  
since they have in common with SFXTs both the ability to trigger the BAT 
and some timing, energetics, and/or spectral properties. 

For instance, the jetted tidal disruption event (TDE) \object{Swift~J164449.3$+$573451} \citep[][]{Burrows2011:NaturJ1644} 
that repeatedly triggered the BAT starting from 2011 March 28 
\citep[][]{Cummings2011:GCN11823,Suzuki2011:GCN11824,Sakamoto2011:GCN11842},  
during the first day since discovery showed characteristics that were reminiscent 
of those of SFXTs \citep[][]{Kennea2011:atel3242}. 
Similarly, the milli-second pulsar (MSP) \object{IGR~J18245$-$2452} \citep[][]{Papitto2013:18245} 
showed an initial soft X--ray light curve similar to a SFXT one \citep[][]{Romano2013:atel4929}.  
In Table~\ref{sfxtcat2:tab:dataswift_weirdtrans} we report the BAT  triggers from these two sources: 
the progressive number of triggers (Col.\ 2), the BAT trigger number  (Col.\ 3),
the BAT trigger significance of detection  (Col.\ 4),              
the trigger UT dates (Col.\ 5 and 6), 
the references to the discovery  and the in-depth outburst analysis (Col.\ 7 and 8), 
and the source type (Col.\ 9).

We also considered representative objects in the
anomalous X-ray pulsars (AXPs) and 
soft gamma repeaters (SGRs) 
class listed at the McGill Online Magnetar 
Catalog\footnote{\href{http://www.physics.mcgill.ca/~pulsar/magnetar/main.html}{http://www.physics.mcgill.ca/$\sim$pulsar/magnetar/main.html.}} 
\citep[][]{Magnetar_cat} that triggered the BAT.  
Table~\ref{sfxtcat2:tab:dataswift_weirdtrans2}  reports, as a few examples, 
the trigger data  of the Galactic magnetar \object{SGR~1935$+$2154}, first discovered by \sw\  
\citep[][]{Cummmings2014:SGR1935}, as well as two prototypical supergiant HMXBs, 
\object{Vela X-1} and \object{4U~1909$+$07} \citep[see, e.g.,][for a recent review]{Kretschmar2019} 
that are known for showing conspicuous flares \citep[see][and references therein]{Walter2015:HMXBs_IGR}.  

Finally, we note that the  outbursts of SFXTs can be easily distinguished from those of most  Be HMXBs and black hole candidates (BHC) 
because the latter two are usually caught while the source flux is still rising and the whole outburst lasts considerably longer, from days to months
(see, e.g., \citealt[][]{Kennea2015:JHEAP}, and \citealt[][and references therein]{Kennea2011:maxij1659,Kennea2021:swiftj011511.0}),  
while SFXT light curves return to pre-outburst levels within a few hours to a day. 
Therefore we shall not discuss  Be HMXBs and BHC flares further.

   \subsection{The {\emph Swift} Burst Analyser \label{sfxtcat2:sba}} 

All  SFXT data  collected after a standard, i.e.\ GRB-like, BAT trigger 
were processed with a minimally\footnote{The SFXT version of the code uses \ XRT hardness ratios different 
from the GRB version, due to the harder spectra of these latter objects.} 
modified version of the \sw\ 
Burst Analyser\footnote{\href{http://www.swift.ac.uk/burst_analyser}{http://www.swift.ac.uk/burst\_analyser.}}  
 \citep{Evans2010:burstanalyser}, whose methods we only summarise here 
while highlighting the special arrangements made for the case of SFXTs, 
whose properties differ from those of GRBs for which it was intended. 

\begin{figure*} 
\vspace{-0.5truecm}
\includegraphics[width=19cm]{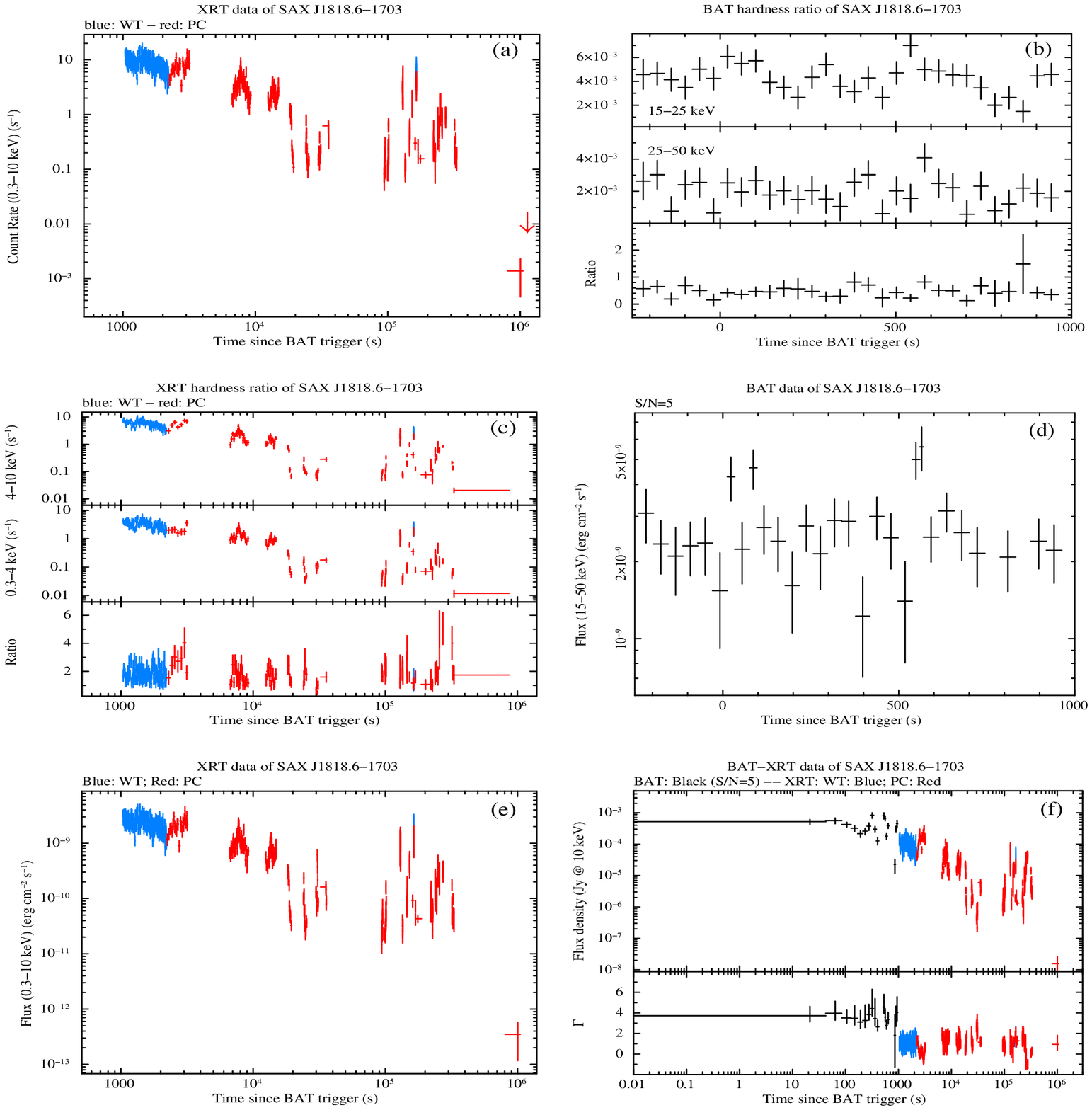}
\vspace{-8truecm}
\caption{Example of Burst Analyser products for the third recorded outburst from SAX~J1818.6$-$1703
(see Table~\ref{sfxtcat2:tab:dataswift}). 
(a): XRT 0.3--10\,keV count rate light curve; 
(b): BAT 15--25\,keV,  25--50\,keV,  and (25--50\,keV/15--25\,keV) hardness ratio; 
(c): XRT 4--10\,keV, 0.3--4\,keV and (4--10\,keV/0.3--4\,keV) hardness ratio; 
(d): BAT 15--50\,keV spectral evolution dependent flux light curve;
(e): XRT 0.3--10\,keV spectral evolution dependent, unabsorbed flux light curve; 
(f): evolution dependent flux density light curves at 10\,keV.} 
\label{sfxtcat2:fig:example_ba} 
\end{figure*}

The \sw\ Burst Analyser uses scripts based on {\sc FTOOLS}\footnote{
\href{https://heasarc.gsfc.nasa.gov/ftools/ftools_menu.html}
{https://heasarc.gsfc.nasa.gov/ftools/ftools\_menu.html.}}  and 
the calibration database CALDB\footnote{\href{https://heasarc.gsfc.nasa.gov/docs/heasarc/caldb/caldb_intro.html}
{https://heasarc.gsfc.nasa.gov/docs/heasarc/caldb/caldb\_intro.html.}} 
to manipulate the event data of GRB BAT triggers collected by BAT 
and XRT and, by using the hardness ratio information in each instrument, 
tracks the spectral evolution, and allows the conversion of the count-rate light curves from both 
 BAT and XRT into accurate, evolution-corrected flux light curves. 
Figure~\ref{sfxtcat2:fig:example_ba} shows an example of the products of the Burst Analyser, 
which end product is the BAT$+$XRT flux density light curves at 10\,keV (Fig.~\ref{sfxtcat2:fig:example_ba}f).

In detail, the BAT data are processed with the tool {\sc batgrbproducts}, which yields information that we can later use 
as diagnostics. 
These include, in keeping with the GRB terminology to define the typical timescales and energetics of the emission,
the trigger time $T_{0}$  (measured in mission elapsed time, MET, i.e., in seconds since 2001-01-01), 
which is the origin of the time ($t=T+T_{0}$, where $t$ is in MET and $T$ is in seconds since the trigger); 
and $T_{90}$, the time during which 90\,\% of the fluence is emitted (from 5\,\% to 95\,\%).  
It also calculates the fluences (in units of on-axis counts per fully illuminated detectors) in the 
standard BAT energy bands,  15--25, 25--50, 50--100\,keV, that can be used to calculate 
hardness ratios and colours \citep[see][for further details]{Evans2010:burstanalyser}. 
Examples of this procedure for SFXTs are the 
BAT 25--50\,keV/15--25\,keV hardness ratios (Fig.~\ref{sfxtcat2:fig:example_ba}b) and the 
15--50\,keV spectral evolution dependent fluxed light curves (Fig.~\ref{sfxtcat2:fig:example_ba}d). 

To convert the count-rate light curves into spectral evolution dependent flux light curves 
a time-evolving counts-to-flux conversion factor is required for  BAT and  XRT,
separately,  and with a single HR per instrument, only one free parameter can be derived. 
As the sources are generally too faint in the BAT band to provide time-resolved
spectroscopy, a spectral shape is assumed and the hardness ratios used to 
track the evolution.  
Therefore a single spectrum is created for all available BAT data and this is fitted 
with both a power-law and cut-off power-law spectral model. 
If the latter gives a $\chi^2$ value of at least 9 lower than the former 
(i.e.\ a 3-$\sigma$ improvement) then the cut-off power-law model is used to 
determine the counts-to-flux conversion factor. 
The cut-off energy is assumed not to vary. 

The XRT data are similarly processed with the tool {\sc xrtpipeline}  and, as described in  full in 
\citet[][]{Evans2007:repository,Evans2009:xrtgrb}, 
light curves are created in the standard XRT energy bands (0.3--1.5, 1.5--10, 0.3--10\,keV for GRBs; 
0.3--4, 4--10, 0.3--10\,keV for SFXTs). 
The  hardness ratio 1.5--10\,keV/0.3--1.5\,keV (4--10\,keV/0.3--4\,keV  for SFXTs) is also created 
(e.g.\ Fig.~\ref{sfxtcat2:fig:example_ba}a and \ref{sfxtcat2:fig:example_ba}c). 

For XRT the Burst Analyser adopts an absorbed power-law, with  
the absorption model consisting of two components, \textsc{phabs} in XSPEC \citep[][]{Arnaud1996:xspec}, 
one fixed at the Galactic column density \citep{LABS}, 
and one to represent the local absorption. 
The absorption is therefore determined by extracting a spectrum from all available 
XRT data corresponding to the trigger, and fitting an absorbed power-law model to 
this spectrum.  As with the cut-off energy, 
this absorption is thereafter assumed to be unchanging as the flare progresses.

For each of BAT and XRT spectra, the spectral model thus determined is loaded into \textsc{XSPEC}
and the photon index of the (cut-off) power-law, $\Gamma$, is varied in the range $-2.5$ to 5. 
A look-up table is thus created (for each $\Gamma$ value)  of 
the hardness ratio, 
the  unabsorbed 0.3--10\,keV flux, 
the  unabsorbed 15--50\,keV flux, 
the model normalisation 
and  the  count-rate predicted by the model, and from those, 
the  hardness ratios vs.\ count-rate-to-flux conversion factors
(Fig.~\ref{sfxtcat2:fig:example_ba}f, bottom panel, reports the photon indices). 

The normalisation of the power-law and cut-off power-law models in
\textsc{XSPEC} is defined as the photon flux density at 1\,keV, in units of photons
cm$^{-2}$\,s$^{-1}$\,keV$^{-1}$. 
In terms of energy flux density normalisation, 
1 keV (cm$^{-2}$\,s$^{-1}$\,keV$^{-1}$) is thus equivalent
to 662.5 $\mu$Jy. This can then be extrapolated to give the flux density
at 10\,keV (Fig.~\ref{sfxtcat2:fig:example_ba}f, top panel). 
By interpolating within the look-up tables, the hardness ratios can be converted 
into time evolution histories of count-rate-to-flux conversion factors. 
Further details can be found in \citet[][]{Evans2010:burstanalyser,Evans2007:repository,Evans2009:xrtgrb}. 
We note that we chose 10\,keV for the flux density, since it is intermediate between the XRT and BAT bands,
hence the extrapolation is minimal.

We note that approximations were made, i.e.\ a constant absorption for the XRT spectra and a 
constant energy of the cut-off for the BAT spectra, but their impact on our results is negligible. 
In Table~\ref{sfxtcat2:tab:dataswift}, we report the references for the in-depth analysis of each dataset, 
specifically tailored to the SFXT case, where the broad-band spectroscopy is performed with several spectral models 
 (in case, with a cut-off energy as a free parameter) and absorption  variations sought for. 
However, here we stress that variable absorbing column densities during a flare were rarely observed in the XRT data 
\citep[e.g.][]{Romano2007,Romano2009:sfxts_paper08408,Sidoli2009:sfxts_paperIV}, 
but in general \citep[e.g.][]{Romano2011:sfxts_paperVI,Romano2014:sfxts_paperX}, 
they are not very strongly supported by the data, if not in rare cases, since the XRT effective area is 
too small to measure fast variations on shorter timescales that can be probed by higher effective area instruments such as
the EPIC cameras on-board \xmm, with which truly remarkable variations have been measured \citep[e.g.][]{Bozzo2017:environ}. 

Discussing the ensemble of the detailed analyses reported in Table~\ref{sfxtcat2:tab:dataswift} is beyond the scope of this paper, 
so we refer the reader to the appropriate references where it was performed, justified by the fact that 
the spectroscopic properties are not an efficient method for discriminating SFXTs within the HMXB population \citep[e.g.][]{Romano2015:swift10}, 
differently from what happens for the dynamic range, as we show below in Sect.~\ref{sfxtcat2:softxray}. 
Indeed, the practice is to fit the SFXT spectra (broad-band or single band) with models that apply to 
NS binaries. Even the models that were specifically developed for SFXTs, such as the 
physical model {\tt compmag} in XSPEC \citep[][]{Farinelli2012:compmag}, 
which includes thermal and bulk Comptonization for cylindrical accretion onto a 
magnetised neutron star, can be successfully applied not only to the SFXT prototypes XTE~J1739$-$302 and 
IGR~J17544$-$2619 \citep[][]{Farinelli2012:sfxts_paperVIII},  and to IGR~J18483$-$0311 \citep[][]{Ducci2013:18483},
but also to several magnetised accreting pulsars \citep[e.g.][]{Farinelli2016:compmag}.

   \subsection{Improved X--ray positions from XRT data \label{sfxtcat2:new_positions}}

For all sources observed by \sw, independently on whether they triggered the BAT or not, 
we derived astrometrically corrected X-ray positions by  
using the XRT-UVOT alignment and matching to the 
USNO-B1 catalogue \citep{Monet2003:USNO_mn} 
as described in 
\citet{Goad2007:xrtuvotpostions} and \citet{Evans2009:xrtgrb}\footnote{
\href{http://www.swift.ac.uk/user_objects/}{http://www.swift.ac.uk/user\_objects/.}}. 
Table~\ref{sfxtcat2:tab:new_positions} reports the cases where this new processing 
produced improved X--ray positions  (significantly smaller 90\,\% confidence error 
circles\footnote{See \href{https://www.swift.ac.uk/user_objects/docs.php}{https://www.swift.ac.uk/user\_objects/docs.php\#posform}.}) 
compared with previous publications in the literature.  
These are currently the best X--ray positions for this subset of SFXTs, although compatible 
to within the uncertainties with previous results.

\setcounter{table}{2}	 
\begin{table}[t]
\tabcolsep 4pt   	
\caption{Improved X--ray positions. \label{sfxtcat2:tab:new_positions}}
\centering
\begin{tabular}{lllrr}
\hline \hline
\noalign{\smallskip} 
Name$^{\mathrm{a}}$              & RA        &Dec                                            &Error  & \\ 
                        & (J2000)  &(J2000)                                       &                             &   \\ 
\hline  \noalign{\smallskip}
IGR~J08408$-$4503   &08 40 47.77 &$-$45 03 30.5   &1\farcs4          &\\  
IGR~J16328$-$4726   &16 32 37.87  &$-$47 23 42.4   &1\farcs4           &\\  
IGR~J16418$-$4532   &16 41 50.90  &$-$45 32 26.7   &1\farcs4           &\\  
IGR~J16465$-$4507    &16 46 35.28   &$-$45 07 05.2  &1\farcs9       &\\  
AX~J1845.0$-$0433   &18 45 01.62   &$-$04 33 56.6  &1\farcs4           &\\   
\noalign{\smallskip}
\hline
\end{tabular}
\tablefoottext{a}{The positions for the full SFXT sample can be found in Table~1 of Paper~I.}
\end{table}

   \section{Results and discussion\label{sfxtcat2:results}}

Currently, most new Galactic transients are either discovered in the deep 
hard X--ray surveys (i.e., \inte/IBIS and \sw/BAT) or as previously unknown or unidentified sources 
that trigger hard X--ray monitors and GRB-chasing missions (such as \sw{} and MAXI). 
Since our ultimate goal is to provide a set of diagnostics to discriminate SFXT candidates from 
newly discovered hard X--ray transients,   we shall first characterise the SFXT `transient' behaviour 
in terms of length of emission, fluences, spectral properties, and 
then compare them with the corresponding ones for GRBs 
and other relevant Galactic and extragalactic transients. 

We shall exploit the observed diagnostics in the chronological order they become available to 
to us, starting from the information distributed through the 
Gamma-ray Coordinates Network\footnote{\href{https://gcn.gsfc.nasa.gov/}{https://gcn.gsfc.nasa.gov/}.} (GCN) 
Notices\footnote{\href{https://www.swift.ac.uk/gcn/index.php}{https://www.swift.ac.uk/gcn/index.php}.} and 
then proceed with information that can only be derived through the analysis of the transient downlinked data. 

   \subsection{SFXTs are image triggers \label{sfxtcat2:image_triggers}}

As we discussed in Paper I, the BAT on-board trigger algorithm 
\citep[][]{Fenimore2003:BATtrigger_algorithm}, works on several different 
timescales and three types of triggers. 
Two are based on increases of count rates 
(short rate triggers on timescales of 4--64\,ms; 
long rate triggers, 64\,ms--24\,s), 
and one is based on images (image triggers, 64\,s to many minutes). 
In the case of image triggers, significant sources are sought in each image, 
and if the source is known and its image flux exceeds a threshold set in the on-board 
source catalog, then \sw\ will slew to it. 
For known SFXTs the on-board threshold was set very low until the end of 2014, 
so that \sw\ would respond with a slew whenever they became significantly brighter. 
The type of trigger (rate vs.\ image) and its duration are in the information distributed with the GCN Notices, 
within seconds to  minutes from the trigger.

All SFXT triggers recorded on board 
are image triggers. The image length in which SFXTs are generally triggering 
varies from 64\,s to 1600\,s, 
the latter being the longest continuous pointing \sw\ can perform, hence the longest timescale available for image 
creation\footnote{Fig.~2 of Paper~I shows the durations of the BAT triggers and a comparison with the
BAT flares registered both on-board and on the ground.}. 
GRBs, on the other hand, based on our analysis of 839 triggers, 
are detected as image triggers only about 13\,\% of the time. 
Both Swift~J164449.3$+$573451 and IGR~J18245$-$2452 produced image triggers,  
lasting from 64 to 1208\,s for the former, and from 64 to 320\,s  for  the latter, 
so that events such as these are not 
easily set apart from the SFXT population based on the triggering method only. 
On the contrary, AXPs/SGRs and magnetar flares are all rate triggers with durations 
from  0.008 to 0.128\,s, hence  they can be set well apart from the SFXT phenomenology 
merely hundreds of seconds after the trigger. 

We also note that the SFXTs IGR~J08408$-$4503, 
IGR~J16479$-$4514, XTE~J1739$-$302, and IGR~J17544$-$2619 
have sometimes triggered the BAT twice in about an hour 
(more often within a few minutes, see Table~\ref{sfxtcat2:tab:dataswift}), an indication that
the source flux is steadily rising. 
While double triggers can occur in jetted TDEs (e.g.\ Swift~J164449.3$+$573451)
and most of all in Galactic transients (see Table~\ref{sfxtcat2:tab:dataswift_weirdtrans}), 
 they are on the other hand very uncommon in GRBs. 
This could be the case of so-called  ultra-long GRBs, which are indeed rare in the 
observed GRB population \citep[][]{Levan2014}.

\begin{figure}[t] 

\vspace{-0.4truecm}

\hspace{+0.25truecm}
\includegraphics[height=9.8cm, angle=-270]{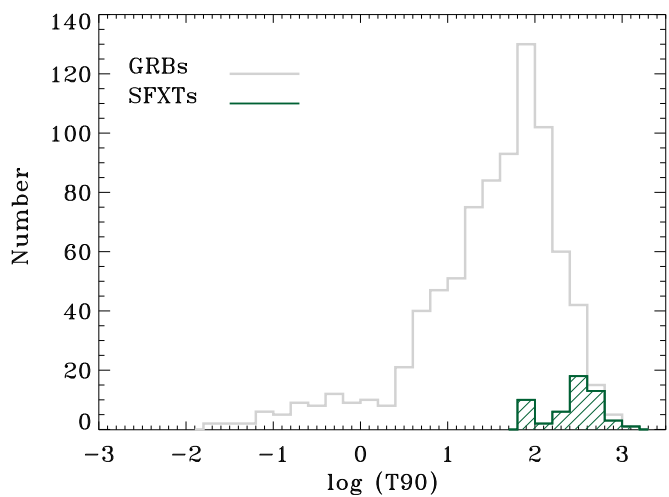}
\caption{Distribution of the $T_{90}$ (s) for the  SFXT triggers ($N=53$, hashed dark green histogram), and for 
the GRB triggers as calculated by the Burst Analyser ($N=839$, light grey line).} 
\label{sfxtcat2:fig:histo_t90spae} 
\end{figure}

\begin{figure*}[h] 
\vspace{-1truecm}

\includegraphics[width=18.5cm]{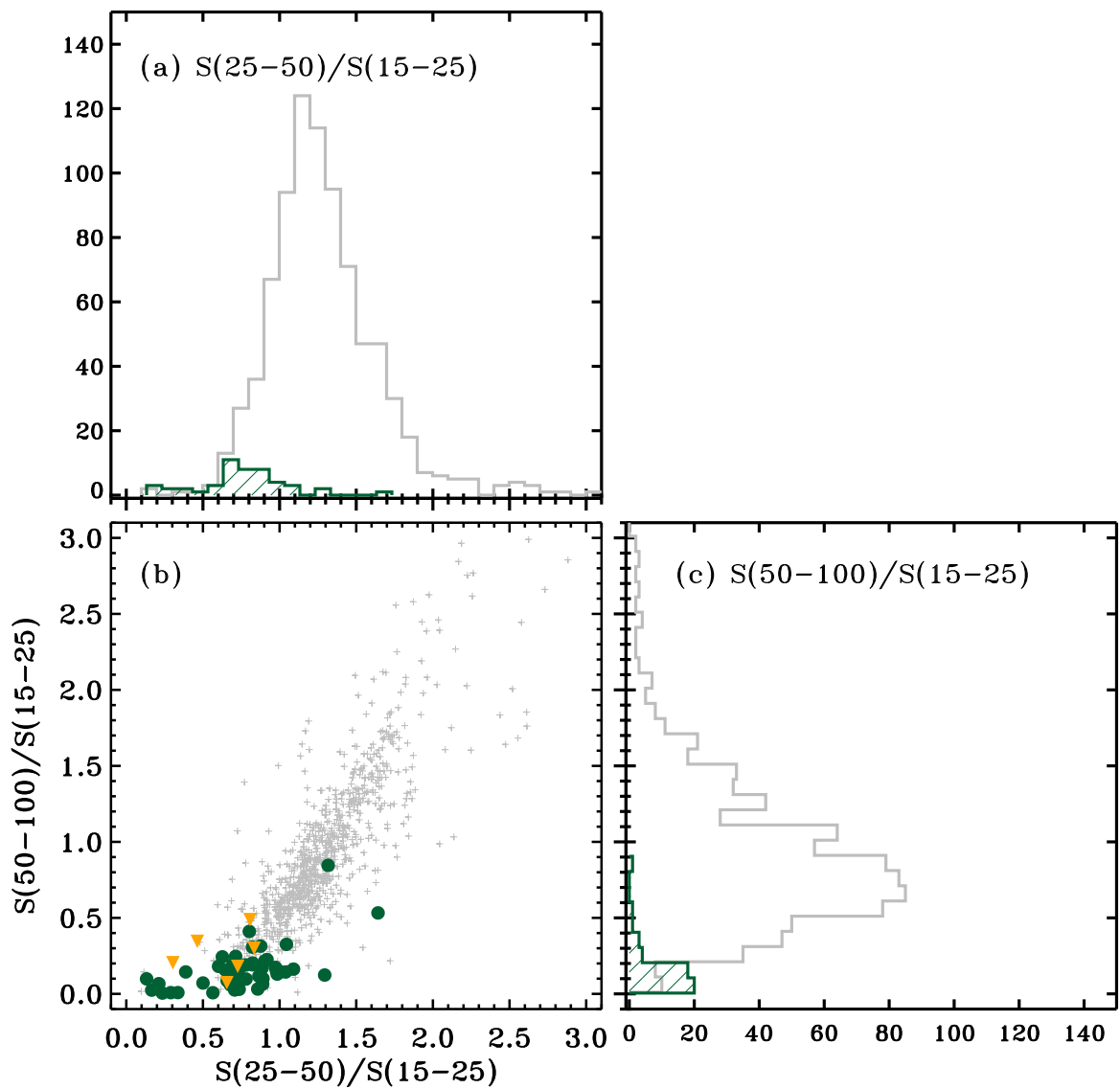}
\vspace{-2.5truecm}
\caption{Comparison of energetics of SFXT ($N=48$, dark green) and GRB ($N=833$, grey):  
(a) distributions of the  $S_{21}=S(25-50)/S(15-25)$ colours;   
(b) colour-colour diagram drawn from the 15--25\,keV, 25--50\,keV, and 50--100\,keV energy bands, 
with the orange triangles ($N=6$) representing 3\,$\sigma$ upper limits for $S(50-100)$;  
(c) distributions of the  $S_{31}=S(50-100)/S(15-25)$  colours.} 
\label{sfxtcat2:fig:fluences_and_colors} 
\end{figure*}

   \subsection{SFXTs are very long transients \label{sfxtcat2:long_grbs}}     

The second diagnostic we consider, 
calculated from the transient downlinked and ground-processed data (hence hours since the trigger) 
is $T_{90}$, which we show in Fig.~\ref{sfxtcat2:fig:histo_t90spae}. 
The grey histogram represents the GRB sample ($N_1=839$) as calculated by the Burst Analyser, 
and shows the two peaks due to the short and long GRB populations \citep[see, e.g.\ ][]{Kouveliotou1993}. 
The dark green hashed histogram shows the values for the SFXT sample, 
also calculated by the Burst Analyser. 
The overall range, based on all usable data  points ($N_{\rm TOT}=53$) is between 64\,s and  1344\,s, and 
the mean is 338$\pm$32\,s. 
Some of these  ($N=30$) are, however, derived from the TDRSS (on-board) values, as the algorithm calculating  $T_{90}$ converged on 
the image duration ({\sc battblocks} failed). The strictly `ground-calculated' sample ($N_2=23$) yields a range of 92--795\,s and a 
mean of 429$\pm$40\,s.

We note that most of the ground-calculated $T_{90}$ for SFXTs are also to be considered realistically lower limits,
since they are based on BAT event mode data only, 
which are telemetered to the ground typically from T$_0-$239\,s to T$_0+$963\,s.  
Indeed, most times BAT registers emission not only well before the trigger but also 
for several hundred seconds after the event list ends, as shown in the survey mode data
collected by the BAT. SFXTs therefore could fall on the high-end tail of the $T_{90}$ distribution, 
and should therefore be compared with long GRBs. 

We performed a Kolmogorov Smirnov test \citep[][]{Darling1957:KS,NR_C_2} 
on the GRB ($N_1=839$) and complete SFXT ($N_2=53$) samples of $T_{90}$  and obtained a K-S statistic $D_{N_1,N_2}= 0.651$ and a K-S probability of $2\times10^{-19}$, so that the two underlying one-dimensional probability distributions differ significantly. Even when excluding all GRBs with $T_{90}<10$\,s (thus effectively excluding short GRBs) $D_{656,53}= 0.622$ and the K-S probability is  $1\times10^{-17}$, thus confirming that the $T_{90}$ of SFXTs and long GRBs are not drawn from the same parent distribution. When considering the strictly ground-calculated sub-sample, we obtain $D_{839,23}= 0.810$ and a probability of $6\times10^{-14}$, while $D_{656,23}= 0.780$ and a probability $7\times10^{-13}$, so our conclusions are not modified by the use of the complete SFXT sample. 

As a comparison, the  jetted TDE Swift~J164449.3$+$573451 and the MSP IGR~J18245$-$2452 have 
$T_{90}$ in the 64--1208\,s and 55--120\,s range, respectively. 
Therefore such events are not distinguishable 
from the SFXT population based on their $T_{90}$ only.  

At odds with SFXTs flares, AXPs/SGRs and magnetar flares have $T_{90}$  
in the 0.008--950\,s range, but most of them 
are below 1\,s. 
Therefore, even when a triggering algorithm did not include image triggers,
AXPs/SGRs and magnetar flares would not be confused with SFXTs, thanks to their
short duration.

   \subsection{SFXTs are faint and `soft'  hard-X--ray transients \label{sfxtcat2:long_xrf}}

\begin{figure} 

\vspace{-0.4truecm}

\hspace{+0.2truecm}
\includegraphics[width=7cm,height=9.9cm, angle=-270]{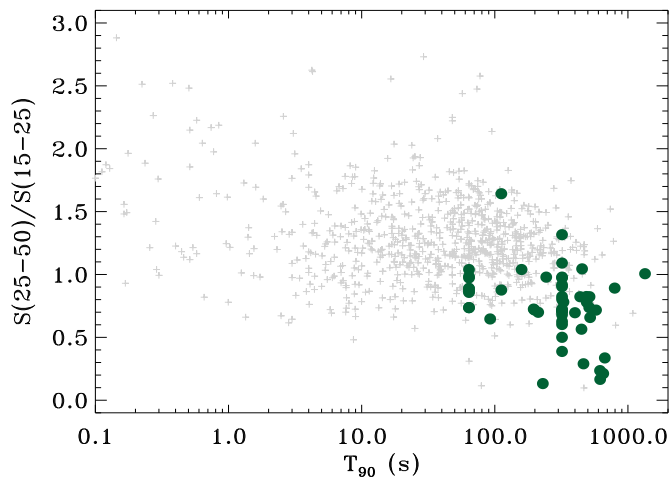}
\caption{Colours $S_{21}=S(25-50)/S(15-25)$  as a function of $T_{90}$. 
Dark green filled circles are the SFXTs, light grey crosses the GRBs.} 
\label{sfxtcat2:fig:t90_color12} 
\end{figure}

\begin{figure}

\vspace{-0.4truecm}

\hspace{+0.2truecm}
\includegraphics[width=7cm,height=9.9cm, angle=-270]{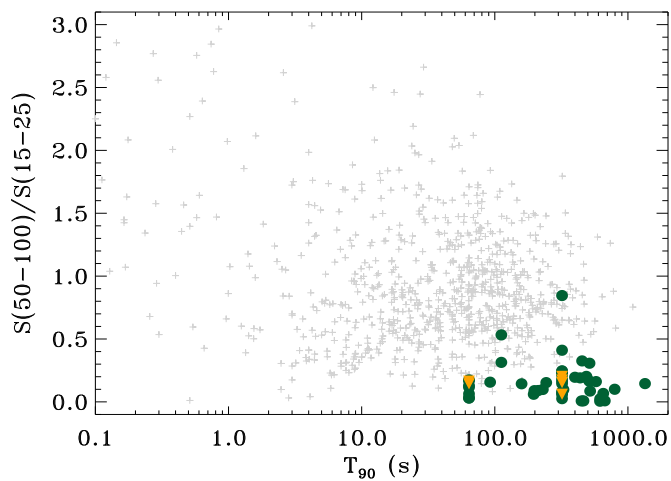}
\caption{Colours $S_{31}=S(50-100)/S(15-25)$ as a function of $T_{90}$. 
Dark green filled circles are the SFXTs, orange triangles  ($N=6$) representing SFXT 3\,$\sigma$ upper limits for $S(50-100)$,
light grey crosses the GRBs.} 
\label{sfxtcat2:fig:t90_color13} 
\end{figure}

\begin{figure}[t]

\vspace{-0.4truecm}

\hspace{+0.2truecm}
\includegraphics[width=7cm,height=9.9cm, angle=-270]{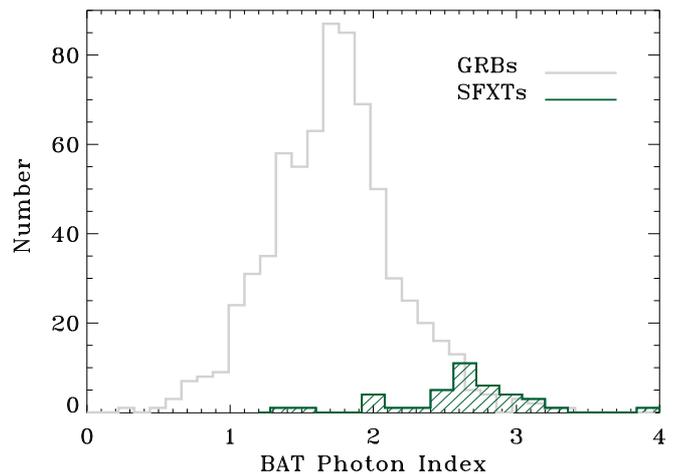}
\caption{Distribution of the photon indices for the SFXT triggers (hashed green histogram) and GRBs
(light grey histogram) when the BAT spectra are fit with a simple power-law model.} 
\label{sfxtcat2:fig:histo_gammas} 
\end{figure}

\begin{figure*} 
\includegraphics[height=23.5truecm,width=19cm]{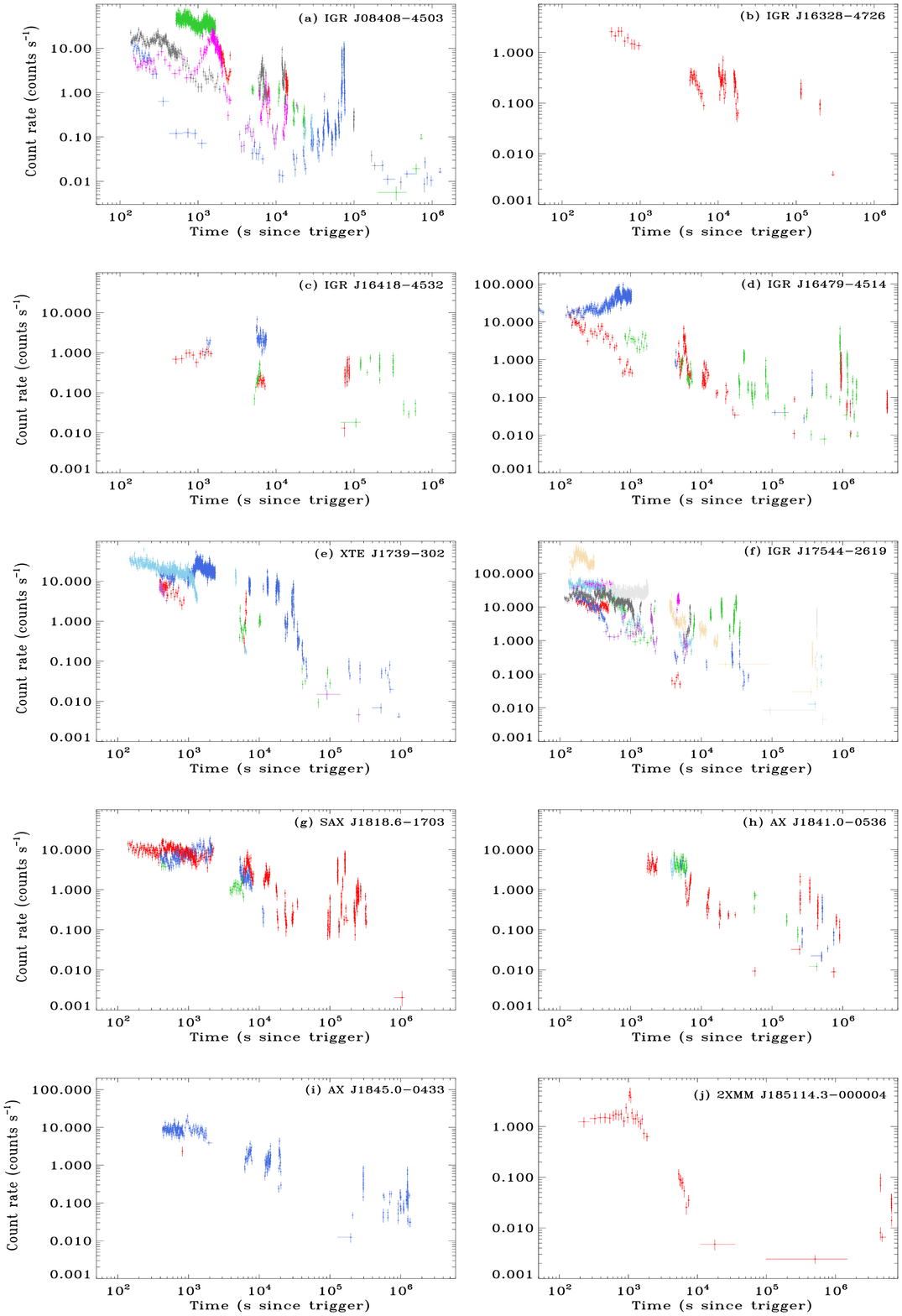}
\hspace{-0.5truecm}

\vspace{-12pt}
\caption{XRT light curves counts s$^{-1}$ in the 0.3--10\,keV energy band, grouped by source 
[(a) IGR~J08408$-$4503, 
(b) IGR~J16328$-$4726, 
(c) IGR~J16418$-$4532, 
(d) IGR~J16479$-$4514, 
(e) XTE~J1739$-$302, 
(f) IGR~J17544$-$2619, 
(g) SAX~J1818.6$-$1703, 
(h) AX~J1841.0$-$0536, 
(i) AX~J1845.0$-$0433, 
(j) 2XMM~J185114.3$-$000004],
as a function of time (s) since the BAT trigger, where different colours correspond to different events. 
The complete list of triggers for each source can be found in Table~\ref{sfxtcat2:tab:dataswift}.
} 
\label{sfxtcat2:fig:xrt_single_lcvs} 
\end{figure*}

The energetics of transients are customarily described by the fluence in several bands, 
which the Burst Analyser calculates in units of on-axis counts per fully illuminated detector. 
We considered the 15--25\,keV (soft), 25--50\,keV (medium), and 50--100\,keV (hard) energy bands,  
measured fluences for both SFXTs and GRBs, and calculated the corresponding colours, 
$S_{21}=S(25-50)/S(15-25)$ and  $S_{31}=S(50-100)/S(15-25)$. 

The average SFXT fluences in the soft band are consistent with those of GRBs, with the 
$S(15-25)_{\rm SFXT}=1.76\pm0.42$\,cts\,det$^{-1}$ and $S(15-25)_{\rm GRB}=1.68\pm0.18$\,cts\,det$^{-1}$,      
but they are significantly fainter than GRBs above 25\,keV, in the medium and hard bands. Indeed, they emit most of their hard X--ray energy in the 15--25\,keV range then, in increasingly lower proportion, in the medium and hard bands similarly to the GRB subset of X-ray flashes \citep[XRF, ][]{Heise2001,Sakamoto2005:XRF_HETE,DAlessio2006:XRF_SAX_HETE,Sakamoto2008:XRF_Swift}. Consequently, the SFXT hardness ratios are systematically larger for GRBs than for SFXTs, 
$S_{\rm 21, GRB}=1.31\pm$0.02 and $S_{\rm 31,  GRB}=0.97\pm$0.02, 
$S_{\rm 21,  SFXT}=0.76\pm$0.04 and $S_{\rm 31,  SFXT}=0.15\pm$0.02, although both are affected by large errors. 

Figure~\ref{sfxtcat2:fig:fluences_and_colors}b shows 
the colour-colour diagram with the GRB sample in grey and the SFXT sample in green; 
the orange triangles represent 3\,$\sigma$ upper limits for $S_{31}$ 
due to some SFXTs not being detected in the 50--100\,keV band. 
Figure~\ref{sfxtcat2:fig:fluences_and_colors}a and \ref{sfxtcat2:fig:fluences_and_colors}c 
show the distributions of the $S_{21}$ and $S_{31}$ colours, respectively.  
A K-S test yields for $S_{21}$ a statistic $D_{833,48}= 0.693$ and a probability of  $4\times10^{-20}$, and for $S_{31}$ a statistic $D_{833,48}= 0.867$ and a probability of  $3\times10^{-31}$, so that the two underlying one-dimensional probability distributions differ significantly. 
As an interesting comparison, Fig.~\ref{sfxtcat2:fig:fluences_and_colors_plus}  shows the corresponding values 
for the TDE Swift J164449.3$+$573451 and the MSP IGRJ 18245$-$2452, 
the magnetar SGR 1935$+$2154, and the two SgHMXBs Vela X$-$1, and 4U 1909$+$07.

In order to further distinguish SFXTs from GRBs, 
in Fig.~\ref{sfxtcat2:fig:t90_color12} and   
\ref{sfxtcat2:fig:t90_color13} 
we plot the fluence ratio $S_{21}$ and $S_{31}$ as a 
function of $T_{90}$, respectively, 
with  grey crosses representing the GRBs,  dark green circles the SFXTs (orange triangles are upper limits). 
Fig.~\ref{sfxtcat2:fig:t90_color13} shows in particular the faintness of SFXTs relative to GRBs 
in the hard band, which is a useful property in terms of discriminating these two classes.

The Burst Analyser performs fits of the BAT event data during the whole available time interval. 
For SFXTs the spectra are best fit with simple power-laws in $42/53$ cases, with photon indices ranging between 1.4 and 4.7, 
with a mean of $\Gamma_{\rm SFXT}=2.77\pm0.11$. 
Such soft indices in the BAT band are customary \citep[e.g.][]{Romano2008:sfxts_paperII} 
confirming the findings of most works based on \inte{} data  \citep[see][and references therein]{sidoli18}. 
The distribution of photon indices is shown in Fig.~\ref{sfxtcat2:fig:histo_gammas} as a hashed dark green histogram. The light grey histogram is the distribution of photon indices of GRBs that are best fit with a simple power law ($N_1=709$). 
The mean is $\Gamma_{\rm GRB}=1.72\pm0.02$. 
A K-S test yields for $\Gamma$ a statistic $D_{709,42}= 0.746$  and a probability of  $2\times10^{-20}$, so that the SFXTs and GRBs photon indices are not drawn from the same parent distribution.
This implies that the BAT spectra of SFXTs are significantly softer than those of GRBs. In particular, we note that SFXT photon indices resemble those typical of X-ray flashes \citep[e.g.,][]{Sakamoto2008:XRF_Swift}. SFXTs are therefore described by relatively faint and soft, long transients.

   \subsection{Soft X-ray prompt and longer-term variability \label{sfxtcat2:softxray}}

\begin{figure}[t] 
\vspace{-0.4truecm}

\includegraphics[width=9.2cm]{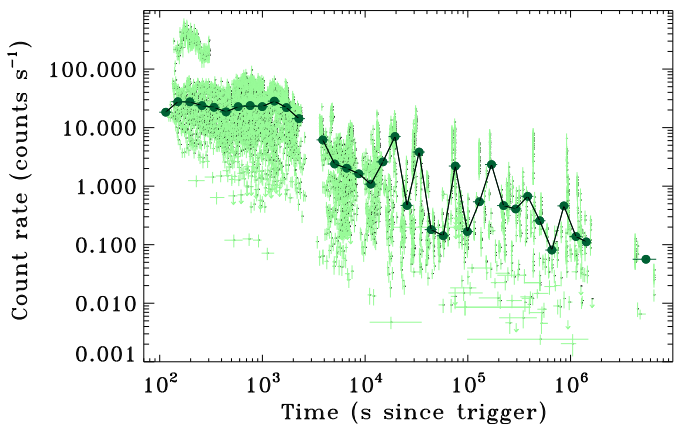}
\caption{XRT count rate light curves (counts s$^{-1}$) in the 0.3--10\,keV energy band 
as a function of time (s since the BAT trigger). 
The dark green points mark the medians. 
} 
\label{sfxtcat2:fig:sfxt_CR} 
\end{figure}

As described in Sect.~\ref{sfxtcat2:sfxtsample}, 35 outbursts (out of 52 events) offer a broad band coverage. 
Here we consider the XRT data for their potential to help us predict the outlook of perspective 
soft X-ray observations following a hard X-ray trigger. 
Figure~\ref{sfxtcat2:fig:xrt_single_lcvs} shows 
for each source the 0.3--10\,keV count rate light curves for each BAT trigger
(see Table~\ref{sfxtcat2:tab:dataswift}) arranged by object\footnote{We note 
that the XRT data of the 2008-03-19 outburst of IGR~J16479$-$4514 start earlier than the BAT trigger, 
since at the time we were monitoring the source regularly \citep[][]{Romano2008:sfxts_paperII}.}  
with the following colour scheme: red, blue, green, light blue, orchid, magenta, light grey. 
Clearly, the light curves are very complex and there is no obvious common behaviour, 
as it varies among different sources and often even among different outbursts of the same source. 
However, there is an overall trend for a fast decay, by up to three orders of magnitude in a few hours.
Re-brightening, in the form of subsequent multiple flares can also occur, as previously reported
\citep[e.g.,][]{Romano2007,Sidoli2008:sfxts_paperI,Romano2009:sfxts_paper08408}. 
We also note that after about 1 day since the trigger there is a flattening of the general trend towards 
the mean value for these sources \citep[see][]{Romano2011:sfxts_paperVI,Romano2014:sfxts_paperX}.  

Figure~\ref{sfxtcat2:fig:sfxt_CR} shows the count rate light curves on the same time scale, 
seconds since the BAT trigger. 
To characterise the overall behaviour, we calculated the median of all light 
curve points in a given time interval, with the time intervals equal in logarithmic space; 
these points (and the connecting line) are plotted in dark green. 
We note that this trend, due to the presence of many orbital and data gaps in the light curves 
and multiple flares, needs to be taken with caution, as it depends on the choice of the time intervals.  

\begin{figure}[t] 
\vspace{-0.4truecm}

\includegraphics[width=9.2cm]{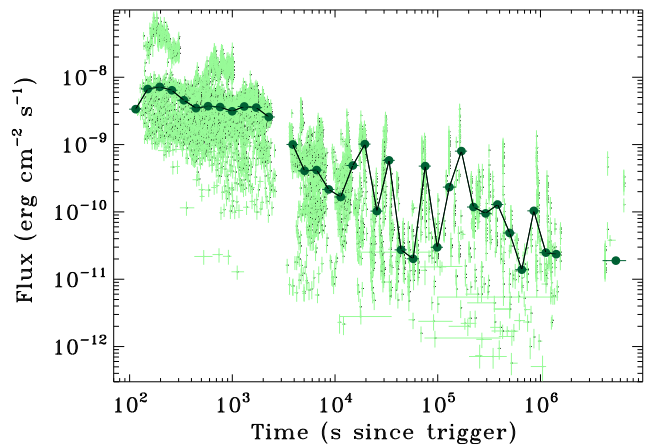}
\caption{XRT  unabsorbed  flux light curves (erg cm$^{-2}$ s$^{-1}$) in the 0.3--10\,keV energy band 
as a function of time (s since the BAT trigger).  
The dark green points mark the medians. 
} 
\label{sfxtcat2:fig:sfxt_flux} 
\end{figure}

Figure~\ref{sfxtcat2:fig:sfxt_flux}, shows the spectral evolution dependent unabsorbed  flux light curves, 
obtained with the methods described in Sect.~\ref{sfxtcat2:sba}, all together.  
 Similarly to what observed in Fig.~\ref{sfxtcat2:fig:sfxt_CR}  
we note that the only shared trait is the lack of a common behaviour with the exception of a 
general fast decay in flux by orders of magnitude in a few hours. 
Although, as stated above, the mean trend needs to be taken with caution, 
we can still exploit its predictive power in terms of the expected flux at a given 
time after the BAT trigger, as is done for GRBs (e.g.\ \citealt[][fig.~11, ]{Evans2009:xrtgrb} and 
\citealt[][Table~3]{Margutti2013:xrtlcv}).

\begin{figure}[t] 

\includegraphics[width=9.2cm]{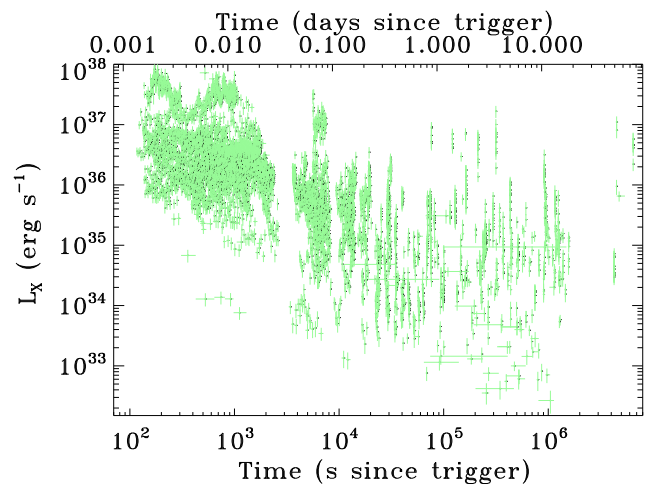}
\caption{XRT unabsorbed  luminosity light curves (erg s$^{-1}$) in the 0.3--10\,keV energy band 
as a function of time (s since the BAT trigger).  
} 
\label{sfxtcat2:fig:sfxt_lum} 
\end{figure}

Finally, Fig.~\ref{sfxtcat2:fig:sfxt_lum} shows the 
unabsorbed   luminosity light curves, 
based on the spectral evolution dependent flux light curves described above and 
the distances reported in Table~\ref{sfxtcat2:tab:sampleproperties}.
 We only propagated the errors on fluxes and not those on distances, the latter being
often just an estimate or missing altogether.

  \setcounter{table}{3}	
 \begin{table*} 	
 \begin{center} 	
  \tabcolsep 3pt
\caption{{\it Swift}/XRT maximum and minimum count rates (counts\,s$^{-1}$) and dynamical ranges, maximum and minimum fluxes (erg\,cm$^{-2}$\,s$^{-1}$) and dynamical ranges.
All are measured in the 0.3--10\,keV energy band from the light curves described in Sect.~\ref{sfxtcat2:softxray}. We also indicated the time after the BAT trigger at which the minimum source flux was recorded.} 	
 \label{sfxtcat2:tab:ranges}
 \begin{tabular}{l ccc ccc  c} 
 \hline 
 \hline 
 \noalign{\smallskip} 
 
  Name                        & CR$_{\rm max}$	 &CR$_{\rm min}$	  	 & CR 	                              &  Flux$_{\rm max}$	              &Flux$_{\rm min}$                   & 	Flux   	                               &  Time$_{\rm min}$  \\
                                   & 	                          & 10$^{-3}$	                 & DR	                                     &  $\times10^{-10}$ 	       & 	$\times10^{-13}$           & DR	                                       &  (d)   \\
 (1) & (2) & (3) & (4) & (5) &  (6) &  (7) &  (8) \\ 
\noalign{\smallskip} 
 \hline 
 \noalign{\smallskip} 
 IGR~J08408$-$4503  &   $73.5\pm11.1$           & $5.6\pm1.9$          & $13012.4\pm4861.2$	 &	$109.3^{+16.6}_{-16.9}$	&	$7.1\pm2.4$	               &	$15348.0\pm5739.9$      &   4.0  	\\
 IGR~J16328$-$4726  & $2.6\pm0.6$               & $<4.6$                   & $573.7$	                         &	$7.7\pm1.7$	                &	$203.1\pm42.3$	       &	$38.2\pm11.7$	         &   0.2  \\
  IGR~J16418$-$4532  & $47.9^{+8.2}_{-8.0}$          & $13.1\pm5.2$      & $3658.3\pm1580.1$	        &	$80.0^{+13.6}_{-13.4}$	&	$120.5^{+26.5}_{-26.5}$	&	$664.1\pm184.9$    &   5.8  	\\
 IGR~J16479$-$4514  & $96.8^{+17.3}_{-14.7}$       & $7.9\pm2.3$        & $12257.8\pm4162.1$       &	$278.1^{+49.7}_{-44.7}$	&	$16.9\pm4.9$	               &	$16490.9\pm5599.8$	&  6.4  \\
 XTE~J1739$-$302     & $61.7\pm9.2$             & $4.6^{+2.1}_{-1.7}$   & $13289.4\pm6332.9$	&	$166.9\pm24.9$	        &	$7.3^{+3.3}_{-2.6}$	       &	$22732.2\pm10832.8$	&  3.0  \\
 IGR~J17544$-$2619  & $604.8^{+130.0}_{-126.2}$    & $4.5^{+1.9}_{-1.5}$    & $133484.1\pm63140.9$	&	$772.6^{+166.1}_{-161.2}$	&	$5.7^{+2.4}_{-1.9}$	      &	$136374.7\pm64508.2$	&  6.0 \\
 SAX~J1818.6$-$1703 &   $19.3\pm2.9$             & $2.0^{+0.9}_{-0.8}$    & $9450.3\pm4500.9$	       &	$43.3\pm6.5$	              &	$5.1^{+2.3}_{-2.0}$          &	$8573.3\pm4087.2$	&  12.1    \\
 AX~J1841.0$-$0536   & $8.4\pm1.5$               & $9.0\pm2.6$          & $938.5\pm317.4$	        &	$14.3\pm3.1$	              &	$17.0\pm4.1$	              &	$840.8\pm271.7$	          &  5.0   \\ 
 AX~J1845.0$-$0433  & $19.8\pm3.4$             & $12.3\pm3.5$        & $1609.2\pm532.3$	       &	$32.1\pm5.5$	              &	$20.0\pm5.7$	              &	$1604.0\pm530.6$	          &   2.3  \\
 2XMM~J185114.3      & $4.8\pm1.1$               & $2.4\pm0.5$         & $1969.6\pm591.9$	       &	$20.6\pm4.6$	              &	$28.0\pm6.7$	              &	$736.7\pm241.6$	          &   0.2  \\   
  \noalign{\smallskip}
  \hline
  \end{tabular}
  \end{center}
  \end{table*}

By using the count rate and spectral evolution dependent flux light curves, 
we can calculate the maximum and minimum in flux and the respective dynamical ranges
in the 0.3--10\,keV energy band for each of the considered SFXTs. 
They are reported in Table~\ref{sfxtcat2:tab:ranges}, where we also indicated the time 
from the BAT trigger at which the source minimum flux was recorded. 

In the soft X-ray domain (0.3--10\,keV), the combination of a fast variability, a several orders of magnitude 
flux decay in just a few hours, and the lack of any major re-brightening within the following few days 
(see Table~\ref{sfxtcat2:tab:ranges}, Fig.~\ref{sfxtcat2:fig:sfxt_CR}, and Fig.~\ref{sfxtcat2:fig:sfxt_lum}), 
remains a key identification criterion for objects in the SFXT class. Excluding GRBs that can be discriminated 
otherwise already by exploiting virtually only BAT data (see previous sections), the other classes of transient 
and variable sources considered in this paper for comparison with the SFXTs show largely different 
behaviours in the XRT energy domain. AXPs/SGRs display much steadier light curves over time 
\citep[see, e.g.,][and references therein]{Esposito2021} and the AMXP IGR~J18245$-$2452 was  
characterised during its outburst by a much slower decrease of the average X-ray flux 
\citep[few days rather than hours;][]{Ferrigno2014}. The peculiar case of the TDE Swift\,J164449.3$+$573451 
showed the necessity of monitoring the transients for at least a few days with XRT to set this event apart 
from an SFXT candidate. The  XRT data collected to within a day from the BAT trigger still showed the TDE to 
closely resemble an SFXT \citep{Kennea2011:atel3242} but on the longer term (a few days) the XRT data 
displayed the presence of multiple major re-flares and re-brightening that do not occur in SFXTs 
\citep[see, e.g.,][]{Burrows2011:NaturJ1644}. The simultaneous availability of the prompt XRT detection 
and monitoring over a few days with the \sw\ NFI following the BAT trigger can therefore provide the 
most solid ground to identify promising SFXT candidates.

   \section{Conclusions \label{sfxtcat2:conclusions}}
In this paper, which complements our previous catalogue of SFXT outbursts detected by BAT 
in the first 100 months of the \sw\ mission \citep[][Paper I]{Romano2014:sfxts_catI}, 
we considered all BAT and XRT data collected as an outburst or as an outburst follow-up. 
Our data-set consists of 52 outbursts (for a total of 56 on-board triggers), of which 
35 have broad band coverage, the great majority of which (27) due to SFXTs being labeled as special sources for the BAT. 
We processed them uniformly with the \sw\ Burst Analyser, and exploited them to derive a set of diagnostics 
that will help distinguish SFXTs triggering future missions scanning the variable X-ray sky with large FoV 
instruments. 
In particular, we concentrated in setting SFXTs apart from the overwhelmingly abundant population of GRBs, 
especially in the early times since the trigger. 
Our systematic investigation of all SFXT outbursts that triggered the \sw/BAT, has led us to the following findings.

\begin{itemize}
\item All SFXT triggers recorded on board BAT are image triggers. 
\item SFXTs are very long transients: the durations range between 64\,s and 1344\,s, with a  mean of $338\pm32$\,s. 
         The $T_{90}$ of SFXTs and long GRBs are not drawn from the same parent distribution.

\item SFXTs are faint hard-X–ray transients:  
most of the SFXT fluence is found in the 15--25\,keV band; 
however, SFXTs are fainter than GRBs in both the 15--25\,keV and 25--50\,keV bands. 

\item SFXTs are ‘soft’ hard-X–ray transients:  
the BAT spectral indices are systematically softer than those of GRBs: 
the mean values are $\Gamma_{\rm SFXT}=2.77\pm0.11$ and  $\Gamma_{\rm GRB}=1.72\pm0.02$; the photon indices of 
SFXTs and GRBs are not drawn from the same parent distribution. 

\item The SFXT properties in the BAT energy band (both fluences and spectral indices) resemble more 
those of X-ray flashes which, however, are generally much shorter transients than SFXTs.  

\item The SFXT X-ray light curves show a common trend for a fast decay, by up to three orders of magnitude, in a few hours
and a flattening after about 1 day after the trigger; 
multiple flaring activity is also observed, leading to a re-brightening up to $\sim$ a few days after the trigger.  
The dynamical ranges in the soft X-rays reach 6 orders of magnitude \citep[IGR~J17544$-$2619, see][]{Romano2015:17544sb}. 
By exploiting the XRT data accumulated as part of the \sw\ triggering observational strategy for SFXTs up to 2014, 
we found that the combination of the SFXT variability in the soft X-ray domain with their fast decay in flux during the first 
few hours after the trigger 
and continuous decay in the following few days is a distinguishable fingerprint for these objects. 
The XRT data can thus be efficiently used to set SFXTs apart not only from GRBs 
(already possible by using virtually only BAT data) but also from other transients such as AXPs/SGRs, 
AMXPs, and jetted TDEs.

\end{itemize}

In summary, we find that SFXTs can be set apart 
from the overall GRB population relatively well already by exploiting BAT trigger data because SFXTs give rise to image triggers 
that are very long, faint, and ‘soft’ in the hard-X–ray domain. The prompt distinction of candidate SFXTs from other classes of transients, such as 
AMXPs and  jetted  TDEs, most likely requires the availability of XRT data from seconds after the triggers up to at least a few days. 
This allows the investigation of the most typical SFXT fingerprint, consisting of a combination of a fast variability in the soft X-ray domain 
with both a hours-long decay in flux by several orders of magnitude after the onset of the outburst and a steady decay in the following days. 

As mentioned earlier in this paper, the threshold for a \sw\ slew for SFXTs was not lowered anymore 
after each outburst after 2014. Consequently, only increasingly brighter outbursts would be 
given a GRB-like response in more recent years, enabling the unique prompt broad band data coverage provided 
by the exploitation of both the BAT and XRT. 
Unless the this strategy changes again, the present data-set is thus unlikely to be significantly extended further.

\begin{acknowledgements}
We thank M.\ Capalbi, D.\ Malesani, and C.\ Ferrigno for helpful discussions. 
We acknowledge unwavering support from Amos. 
We also thank the anonymous referee for comments that helped improve the paper. 
We acknowledge financial contribution from contracts ASI-INAF I/004/11/0,  
ASI-INAF I/037/12/0,  
and ASI-INAF n. 2017-14-H.0,  
HAK acknowledges support while serving at the National Science Foundation. 
The data underlying this article are publicly available from the \sw\ archive.  
This work made use of data supplied by the UK Swift Science Data Centre at the University of Leicester.
The \sw/BAT transient monitor results are provided by the \sw/BAT team. 
Happy 18th, \sw. 
\end{acknowledgements}

\begin{appendix}
\section[T]{Supplementary tables and figures}

\setcounter{table}{0} 
\begin{table*}
\tabcolsep 6pt         
\caption{\sw\ data for the TDE Swift J164449.3$+$573451 the MSP IGRJ 18245$-$2452.\label{sfxtcat2:tab:dataswift_weirdtrans} }
\centering
\begin{tabular}{llcrll ccc}
\hline 
\hline 
\noalign{\smallskip} Name              & \multicolumn{5}{c}{Trigger}                              &\multicolumn{2}{c}{References}  &  Type of \\
                       & N\tablefootmark{a}  & \#\tablefootmark{b}    &    {\it S/N}\tablefootmark{c}         &UT Date            &UT Time         &Discovery   & Refereed   &  Source \\
(1) & (2) & (3) & (4) & (5) &  (6) &  (7) &  (8) & (9)\\ 
\noalign{\smallskip}
\hline
\noalign{\smallskip} 
Swift~J164449.3$+$573451  &1 &450158  &7.60   & 2011-03-28  &12:57:45   &\getrefnumber{tab2Cummings2011:GCN11823},\getrefnumber{tab2Kennea2011:atel3242}  & \getrefnumber{tab2Burrows2011:NaturJ1644}  &  TDE \\  
                                              &2 &450161  &6.57   & 2011-03-28  &13:40:41   &\getrefnumber{tab2Suzuki2011:GCN11824},\getrefnumber{tab2Kennea2011:atel3242}         & \getrefnumber{tab2Burrows2011:NaturJ1644}      \\    
                                              &3 &450257  &10.73 & 2011-03-29  &18:26:25   & \getrefnumber{tab2Sakamoto2011:GCN11842}                                                                      &  \getrefnumber{tab2Burrows2011:NaturJ1644}     &  \\   
                                              &4 &450258  &11.85 & 2011-03-29  &19:57:45   & \getrefnumber{tab2Sakamoto2011:GCN11842}                                                                      & \getrefnumber{tab2Burrows2011:NaturJ1644}      &    \\  
 \noalign{\smallskip}
 \hline
\noalign{\smallskip}
IGRJ~18245$-$2452             &1&552336  &5.86 & 2013-03-30   &02:22:21 & \getrefnumber{tab2Barthelmy2013:GCN14355},\getrefnumber{tab2Romano2013:atel4929} &  \getrefnumber{tab2Papitto2013:18245}   &  MSP \\ 
                                            &2&552369  &8.10 & 2013-03-30   &15:10:37 &                                                                                                                                                 & \getrefnumber{tab2Papitto2013:18245}   &  \\   
\noalign{\smallskip}
 \hline
\noalign{\smallskip} 

\end{tabular}
\tablefoot{
\tablefoottext{a}{Progressive number of BAT trigger.}
\tablefoottext{b}{BAT Trigger number. }
\tablefoottext{c}{On-board significance of detections of BAT trigger in units of $\sigma$.}}
\tablebib{
\newcounter{ctr_apprefs} 
\setrefcountdefault{-999} 
\refstepcounter{ctr_apprefs}\label{tab2Cummings2011:GCN11823}(\getrefnumber{tab2Cummings2011:GCN11823})  \citet{Cummings2011:GCN11823};
\refstepcounter{ctr_apprefs}\label{tab2Kennea2011:atel3242}(\getrefnumber{tab2Kennea2011:atel3242})  \citet{Kennea2011:atel3242};
\refstepcounter{ctr_apprefs}\label{tab2Burrows2011:NaturJ1644}(\getrefnumber{tab2Burrows2011:NaturJ1644})  \citet{Burrows2011:NaturJ1644};
\refstepcounter{ctr_apprefs}\label{tab2Suzuki2011:GCN11824}(\getrefnumber{tab2Suzuki2011:GCN11824})  \citet{Suzuki2011:GCN11824};
\refstepcounter{ctr_apprefs}\label{tab2Sakamoto2011:GCN11842}(\getrefnumber{tab2Sakamoto2011:GCN11842})  \citet{Sakamoto2011:GCN11842};
\refstepcounter{ctr_apprefs}\label{tab2Barthelmy2013:GCN14355}(\getrefnumber{tab2Barthelmy2013:GCN14355})  \citet{Barthelmy2013:GCN14355};
\refstepcounter{ctr_apprefs}\label{tab2Romano2013:atel4929}(\getrefnumber{tab2Romano2013:atel4929})  \citet{Romano2013:atel4929};
\refstepcounter{ctr_apprefs}\label{tab2Papitto2013:18245}(\getrefnumber{tab2Papitto2013:18245})  \citet{Papitto2013:18245}. 
}
\end{table*}

\setcounter{table}{1} 
\begin{table*}
\tabcolsep 3pt         
\caption{\sw\ data for other transients.\label{sfxtcat2:tab:dataswift_weirdtrans2}  }
\centering
\begin{tabular}{ l crrll  crrll   }
\hline 
\hline 
\noalign{\smallskip} Name              & \multicolumn{5}{c}{Trigger}       & \multicolumn{5}{c}{Trigger}                               \\
                       & N\tablefootmark{a}  & \#\tablefootmark{b}    &    {\it S/N}\tablefootmark{c}         &UT Date            & UT Time    & \hspace{1cm} N\tablefootmark{a}  & \#\tablefootmark{b}    &    {\it S/N}\tablefootmark{c}         & UT Date            & UT Time         \\
\noalign{\smallskip}
\hline
\noalign{\smallskip} 
SGR 1935$+$2154 & 1   &  603488 &  14.89 & 2014-07-05 &  09:32:49      & \hspace{1cm} 9  &  701590 &  26.48 &   2016-06-26 &  13:54:31    \\     
  (magnetar )       &  2  &  632159 &  16.47 &  2015-02-22 &  12:31:11      &\hspace{1cm} 10 &  933083 &   11.11&   2019-11-04 &  06:34:00     \\ 
                           &  3  &  686443 &  9.21   &  2016-05-16 &  20:49:47     & \hspace{1cm} 11 &  933276 &   15.03&  2019-11-05 &  00:08:58       \\ 
                           &  4  &  686761 &  24.16&  2016-05-18  &  09:09:24      & \hspace{1cm} 12 &  933285 &  23.89&   2019-11-05 &  01:36:26     \\  
                           &  5  &  686842 &  7.01  &  2016-05-19 &  05:41:26       & \hspace{1cm} 13 &  968211 &  19.26 &   2020-04-27 &  18:26:20 \\
                           &  6  &  687123 &  6.93  &  2016-05-21 &  20:01:47        & \hspace{1cm} 14 &  968212 & 19.17 &   2020-04-27 &  18:33:00  \\
                           &  7  &  687124 &  14.02 &  2016-05-21 &  20:23:42       &\hspace{1cm}  15 & 1108538 & 22.93&   2022-05-30 &  20:32:27  \\
                           &  8  &  701182 &  21.52 &  2016-06-23 &  19:24:40       & \\
  \noalign{\smallskip}
\hline
\noalign{\smallskip} 
Vela X-1             & 1  &  500100 & 16.84  & 2011-08-12   &        06:38:04 &  \hspace{1cm}  3   & 618634  & 16.14 &  2014-11-14   &        21:11:12 \\  
  (SgHMXB)	    & 2   & 579975  &12.64  & 2013-12-02   &        12:40:41 & \hspace{1cm}  4   & 779762  & 25.6   & 2017-10-17    &       17:38:10 \\    
 
\noalign{\smallskip}
\hline
\noalign{\smallskip}  
4U 1909$+$07 &1 &156561 &5.57   & 2005-09-23 & 02:12:48  & \hspace{1cm}  3 &292965 & 8.26  & 2007-10-03 & 14:04:46 \\ 
 (SgHMXB)	&2 &157253 &5.63   & 2005-09-27 & 07:02:56  &  \\ 

 \noalign{\smallskip}
\hline
\end{tabular}
\tablefoot{
\tablefoottext{a}{Progressive number of BAT trigger.}
\tablefoottext{b}{BAT Trigger number. }
\tablefoottext{c}{On-board significance of detections of BAT trigger in units of $\sigma$.}}
\end{table*}

\begin{figure*}[h] 
\vspace{-1truecm}

\includegraphics[width=18.5cm]{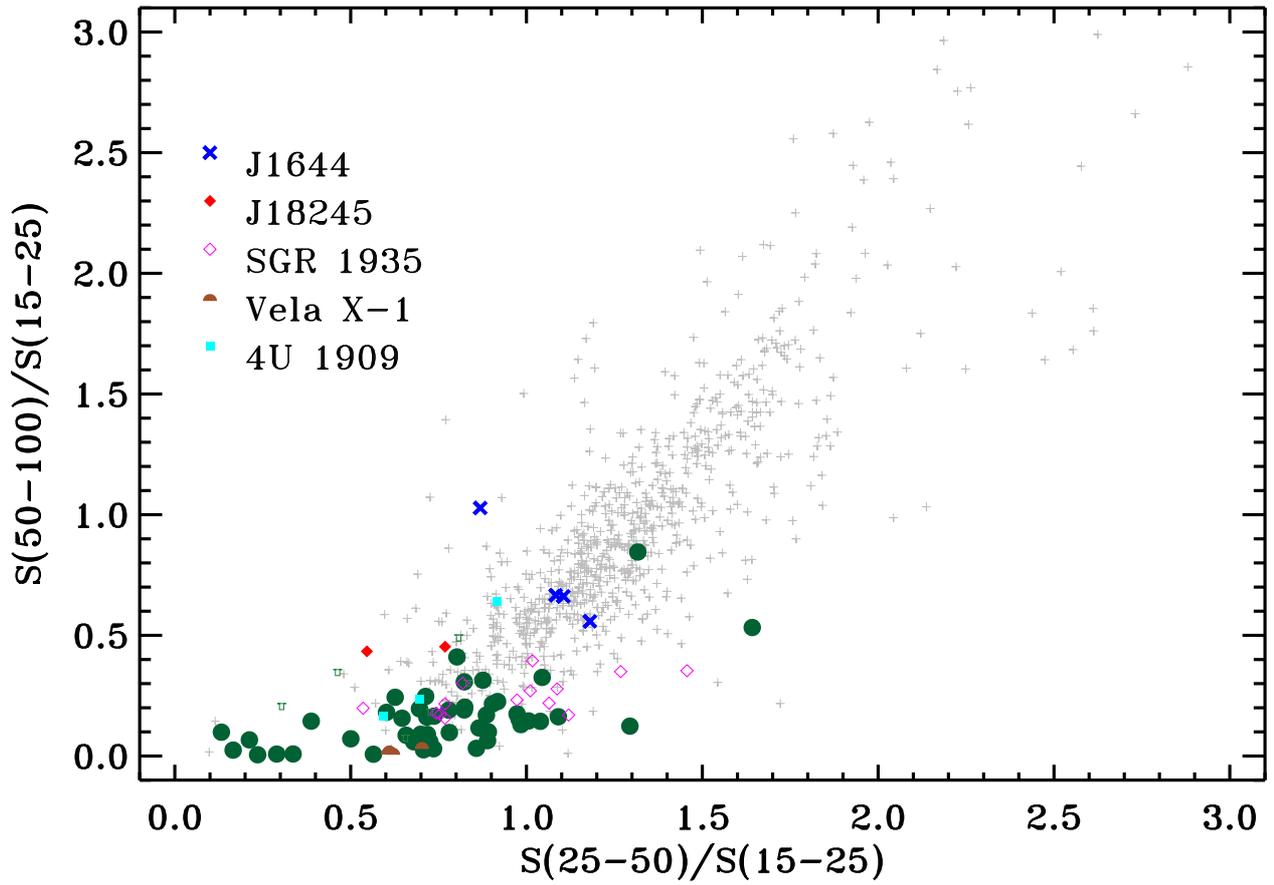}
\caption{Comparison of the energetics of  GRB (grey), 
SFXT (dark green filled circles, with green empty hats representing 3\,$\sigma$ upper limits)  
and selected other transients: the TDE Swift J164449.3$+$573451 (J1644, blue X), 
the MSP IGRJ 18245$-$2452 (J18245, filled red diamonds), see  Table~\ref{sfxtcat2:tab:dataswift_weirdtrans}, 
the magnetar SGR 1935$+$2154 (SGR1935, empty magenta diamonds), 
and the two supergiant HMXBs Vela X-1 (filled upper half circles)
 and 4U~1909$+$07 (J1909, filled cyan squares), see  Table~\ref{sfxtcat2:tab:dataswift_weirdtrans2}.  
\label{sfxtcat2:fig:fluences_and_colors_plus} }
\end{figure*}

\end{appendix}

\end{document}